\documentclass[aps, prl, twocolumn, superscriptaddress, letterpaper]{revtex4-2}

\usepackage{graphicx}
\usepackage{amssymb}
\usepackage{amsmath}
\usepackage{float}
\usepackage{txfonts}
\usepackage{bm}
\usepackage{color}
\usepackage[colorlinks=true, linkcolor=blue, anchorcolor=black, citecolor=blue, filecolor=black, menucolor=black, runcolor=black, urlcolor=blue]{hyperref}
\usepackage{mathrsfs}

\begin{document}

\title{Observation of Embedded Topology in a Trivial Bulk via Projective Crystal Symmetry}

\author{Hau Tian Teo}
\affiliation{Division of Physics and Applied Physics, School of Physical and Mathematical Sciences, Nanyang Technological University, Singapore 637371, Singapore}

\author{Yang Long}
\email{yang.long.physics@outlook.com}
\affiliation{Division of Physics and Applied Physics, School of Physical and Mathematical Sciences, Nanyang Technological University, Singapore 637371, Singapore}

\author{Hong-yu Zou}
\affiliation{Research Center of Fluid Machinery Engineering and Technology,
School of Physics and Electronic Engineering, Jiangsu University, Zhenjiang 212013, China}

\author{Kailin Song}
\affiliation{Division of Physics and Applied Physics, School of Physical and Mathematical Sciences, Nanyang Technological University, Singapore 637371, Singapore}

\author{Haoran Xue}
\affiliation{Department of Physics, The Chinese University of Hong Kong, Shatin, Hong Kong SAR, China}

\author{Yong Ge}
\affiliation{Research Center of Fluid Machinery Engineering and Technology,
School of Physics and Electronic Engineering, Jiangsu University, Zhenjiang 212013, China}

\author{Shou-qi Yuan}
\affiliation{Research Center of Fluid Machinery Engineering and Technology,
School of Physics and Electronic Engineering, Jiangsu University, Zhenjiang 212013, China}

\author{Hong-xiang Sun}
\email{jsdxshx@ujs.edu.cn}
\affiliation{Research Center of Fluid Machinery Engineering and Technology,
School of Physics and Electronic Engineering, Jiangsu University, Zhenjiang 212013, China}
\affiliation{State Key Laboratory of Acoustics, Institute of Acoustics, Chinese Academy of Sciences, Beijing 100190, China}

\author{Baile Zhang}
\email{blzhang@ntu.edu.sg}
\affiliation{Division of Physics and Applied Physics, School of Physical and Mathematical Sciences, Nanyang Technological University, Singapore 637371, Singapore}
\affiliation{Centre for Disruptive Photonic Technologies, Nanyang Technological University, Singapore 637371, Singapore}

\date{\today}

\begin{abstract}
Bulk-boundary correspondence is the foundational principle of topological physics, first established in the quantum Hall effect, where a $D$-dimensional topologically nontrivial bulk gives rise to $(D-1)$-dimensional boundary states. The advent of higher-order topology has generalized this principle to a hierarchical chain, enabling topological states to appear at $(D-2)$ or even lower-dimensional boundaries. To date, all known realizations of topological systems must require a topologically nontrivial bulk to initiate the chain of action for bulk-boundary correspondence. Here, in an acoustic crystal platform, we experimentally demonstrate an exception to this paradigm—embedded topology in a trivial bulk—where the bulk-boundary correspondence originates from a trivial bulk. Rather than relying on global symmetries, we employ projective crystal symmetry, which induces nontrivial topology not at the outset in the $D$-dimensional bulk, but midway through the correspondence hierarchy in lower-dimensional boundaries. We further realize a three-dimensional system exhibiting embedded topology that supports zero-dimensional topological states, achieving the longest possible chain of action for such an unconventional bulk-boundary correspondence in physical space.
Our work experimentally establishes a new form of bulk-boundary correspondence initiated from a trivial bulk, opening additional degrees of freedom for the design of robust topological devices.
\end{abstract}
                    
\maketitle

\begin{figure}[b]
    \centering
    \includegraphics[width=\columnwidth]{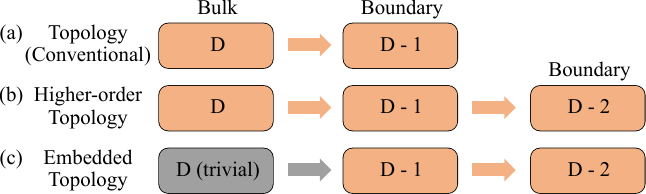}
    \caption{\textbf{Chain of action for the principle of bulk-boundary correspondence in different topology mechanisms.} (a) Conventional topological insulators (i.e., first-order topology), stemming from $D$-dimensional bulk topology, exhibit topological ($D-1$)-dimensional boundary states. (b) Higher-order topological insulators, similarly initiated in \textit{$D$}-dimensional bulk topology, manifest topological boundary states in ($D-2$) dimensions or lower. (c) An embedded topological insulator can host a $D$-dimensional trivial bulk. Nontrivial topology can arise from its ($D-1$)-dimensional boundary, leading to ($D-2$)-dimensional topological boundary states.}
    \label{fig1}
\end{figure}

As the foundational principle of topological physics, bulk-boundary correspondence establishes the corresponding relationship between the global band topology of a system's bulk and the existence of protected states at its boundaries. In its conventional form~\cite{Klitzing1980, Halperin1982, Hasan2010, Bansil2016, Chiu2016, Wang2009, Lu2014, Yang2015}, as exemplified by the quantum Hall effect~\cite{Klitzing1980, Halperin1982}, a $D$-dimensional topologically nontrivial bulk gives rise to $(D-1)$-dimensional boundary states [Fig.~\ref{fig1}(a)].
This framework has been generalized by the concept of higher-order topology~\cite{Benalcazar2017, Benalcazar2017a, SerraGarcia2018, Peterson2018, Mittal2019, Qi2020, He2020, Xue2020, Ni2020, Xie2021}, where the bulk-boundary correspondence becomes hierarchical, enabling topological states to appear at $(D-2)$ or even lower-dimensional boundaries under suitable crystalline symmetries [Fig.~\ref{fig1}(b)]. Yet, across all these frameworks, a prevailing assumption remains: the chain of bulk-boundary correspondence must always be initiated by a bulk that is itself topologically nontrivial. Recent theoretical proposals have begun to challenge this view by introducing the concept of embedded topology~\cite{Tuegel2019,Velury2022,Panigrahi2022}, where the bulk can remain trivial, yet the boundary bands themselves acquire nontrivial topology, leading to lower-dimensional topological boundary states [Fig.~\ref{fig1}(c)]. This unconventional chain for bulk-boundary correspondence broadens the understanding of how symmetries and dimensional hierarchies enable topological phenomena, but it has never been experimentally validated. Moreover, while embedded topology in a trivial bulk has been discussed theoretically under global symmetries, its potential manifestation via crystalline symmetries remains unclear.

Recently, arising from time-reversal-invariant gauge fields within crystal symmetry, projective crystal symmetry has exhibited new topological phases, including M\"obius-twisted edge bands~\cite{Zhao2020, Xue2022, Li2022}, Stiefel-Whitney topological charges~\cite{Shao2021, Xue2023}, and Klein bottle in the Brillouin zone~\cite{Chen2022, Pu2023, Li2023, Tao2024, Zhu2024}.
The topological classification can also be enriched under projective crystal symmetry, with dimension still playing a crucial role~\cite{Zhao2021,Huang2022, Chen2023}. 
However, so far, there is neither theoretical nor experimental work discussing the embedded topology based on the projective crystal symmetry, particularly from the perspective of bulk-boundary correspondence. 

In this Letter, by employing projective crystal symmetry in an acoustic crystal, we report on the first experimental realization of embedded topology in a trivial bulk.
Firstly, we demonstrate a one-dimensional (1D) strong topological insulator (TI) by forming an interface between two-dimensional (2D) trivial insulators.
The 1D strong TI arises from the topological phase transition in the 1D gapless interface, thereby unveiling the embedded topology.
To observe the topological phase of the gapped interface, we connect two 1D topologically distinct interfaces to form a 2D trijunction where an in-gap topological state emerges.
Furthermore, we extend the hierarchy of embedded topology to three-dimensional (3D) insulators, where the 3D bulk and 2D surfaces remain trivial; the embedded topology emerges on 1D hinges and induces a zero-dimensional (0D) topological state at a 3D quadrijunction, thus demonstrating the longest possible hierarchy in physical space.

\begin{figure}[tp!]
  \centering
  \includegraphics[width=\columnwidth]{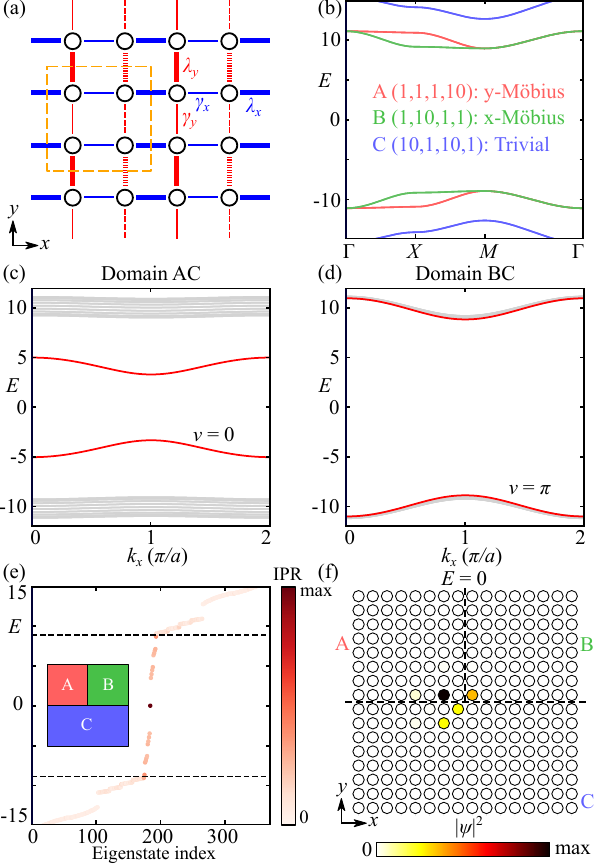}
  \caption{\textbf{Embedded topology under projective crystal symmetry}. (a) Illustration of the 2D tight-binding model. The solid (dotted) lines denote positive (negative) couplings, with their thickness and color indicating different magnitudes $\gamma_{x,y}$ and $\lambda_{x,y}$. The dashed box denotes the unit cell.
(b) Bulk dispersions of insulators A, B, and C with coupling patterns $(\gamma_x, \lambda_x, \gamma_y, \lambda_y)$.
(c)-(d) Edge dispersions of different domains which are periodic along $x$ direction. 
The bulk bands are indicated in grey, while the interface bands are highlighted in red, with their corresponding Zak phase $\nu$ labeled. 
(e)~Eigenenergy spectrum of the finite 2D lattice with a trijunction at the center, color-coded by IPR of each eigenstate. The dashed lines denote the common bulk bandgap.
(f)~Field distribution of the trijunction-induced topological state. The domain walls are represented by the dashed lines.}
  \label{fig2}
\end{figure}

We begin with the 2D tight-binding model in Fig.~\ref{fig2}(a). 
It consists of positive and negative dimerized couplings along $x$ and $y$ directions, where $\gamma_{x,y}$ ($\lambda_{x,y}$) denote the intracell (intercell) couplings.  The bulk Hamiltonian reads: $H_\text{2D} = (\gamma_x + \lambda_x \cos{k_x})\,\Gamma_1 + \lambda_x\sin{k_x}\, \Gamma_2 \nonumber 
	 + (\gamma_y + \lambda_y \cos{k_y})\,\Gamma_3 + \lambda_y \sin{k_y} \,\Gamma_4 $, where $\Gamma_1 = \tau_0\sigma_1$, $\Gamma_2 = \tau_0\sigma_2$, $\Gamma_3 = \tau_1\sigma_3$ and $\Gamma_4=\tau_2\sigma_3$ are constructed by two sets of Pauli matrices $\{\tau_i\}$ and $\{\sigma_i\}$, satisfying $\{\Gamma_\mu, \Gamma_\nu\} = 2\delta_{\mu\nu}\mathbb{I}_4$ and $\gamma_{x,y}, \lambda_{x,y}>0$. 
The topological properties of $H_\text{2D}$ rely on symmetry conditions, e.g., quadrupole $\mathbb{Z}_2$ TI phase if four-fold rotation symmetry is preserved~\cite{Benalcazar2017a, Benalcazar2017, SerraGarcia2018, Peterson2018, Mittal2019, Qi2020, He2020}, and M\"obius insulator phase under projective translation symmetry~\cite{Zhao2020,Xue2022,Li2022}. 
In our work, we only preserve time-reversal symmetry $\mathcal{T}$ and projective mirror symmetry $\mathcal{M}_x = \mathcal{G} M_x$ ($\mathcal{T}^2=\mathcal{M}_x^2 = 1$, $[\mathcal{M}_x, \mathcal{T}] = 0$), i.e.,
\begin{equation}
\mathcal{M}_x H_\text{2D}(k_x,k_y) \mathcal{M}_x^{-1} = H_\text{2D}(-k_x,k_y),
\end{equation}
where $M_x=\tau_0 \sigma_1$ is the mirror operator along $x$-axis up to a gauge $\mathcal{G} = \tau_3\sigma_0$~\cite{Chen2023}.
According to the topological classifications~\cite{Chiu2016, Chiu2013}, 2D insulators under $\mathcal{M}_x$ and $\mathcal{T}$ are topologically trivial.

Note that there can be accidental projective translation symmetry in the model in Fig.~\ref{fig2}(a). 
The accidental projective translation symmetry can be described by the projective translation operators $\mathcal{L}_{x,y}$ with a $\pi$ gauge flux~\cite{Zhao2020,Xue2022,Li2022}. 
By tuning $\gamma_{x,y}$ and $\lambda_{x,y}$, we can have three kinds of dimerizations that break $\mathcal{L}_x$, $\mathcal{L}_y$, or both, corresponding to the insulators A, B, and C in Fig.~\ref{fig2}(b). 
While preserving $\mathcal{L}_x$ or $\mathcal{L}_y$, insulators A and B are in a M\"obius insulator phase~\cite{Zhao2020, Xue2022, Note1}. 
Consequently, while truncating the lattice along $y$ or $x$ direction, a M\"obius twist of edge bands emerges in the bulk bandgap and exists on the $\mathcal{L}_x$- or $\mathcal{L}_y$-preserved edge~\cite{Xue2022,Li2022}. 

By connecting the 2D trivial insulators, the accidental projective translation symmetry is lifted, since there is no common projective translation symmetry among insulators A, B, and C.   
For instance, insulator A with $\mathcal{L}_x$ can be connected to insulator B or C without $\mathcal{L}_x$. 
Thus, the degeneracy in M\"obius twist of edge bands protected by $\mathcal{L}_x$ or $\mathcal{L}_y$ is gapped out.  
Here, we name the interface formed by insulators A and C as domain AC, and other interfaces follow the same naming convention. 
As shown in Fig.~\ref{fig2}(c), we obtain gapped edge bands existing on the interfaces of different domains. 
The interface can be described by a 1D Hamiltonian $H_s$. 
$H_s$ still possesses the projective mirror symmetry $\mathcal{M}_x$: $\mathcal{M}_x H_s(k_x) \mathcal{M}_x^{-1} =  H_s(-k_x)$ and time-reversal symmetry $\mathcal{T}$. 
Since $\mathcal{T}^2=1$ and $[\mathcal{M}_x, \mathcal{T}]=0$, $H_s$ in Figs.~\ref{fig2}(c)-\ref{fig2}(d) can be topological~\cite{Chiu2013, Chiu2016}.
The topological phase is described by an integer invariant $N_- = N_-^\pi-N_-^0 \in~\mathbb{Z}$ (i.e., $M\mathbb{Z}$ topology in class AI), where $N_-^{0}$ ($N_-^{\pi}$) is the number of states with $\mathcal{M}_x$ eigenvalue $-1$ at $k_x=0$ ($k_x=\pi$). 
$N_-$ is also related to the Zak phase $\nu$ of the edge bands: $e^{i\nu} = (-1)^{N_-}$, where $\nu = \oint dk_x\, \mathcal{A}_x \, (\text{mod}\, 2\pi)$ and $\mathcal{A}_x$ is the Berry connection~\footnote{See Supplementary Material for details}.  
$\nu$ is labeled for each interface band below zero energy, as shown in Figs.~\ref{fig2}(c)-\ref{fig2}(d). 
The non-zero $N_{-}$  reflects that $H_s$ describes a 1D strong TI protected by $\mathcal{M}_x$ and $\mathcal{T}$, thus exhibiting the embedded topology protected by the projective crystal symmetry. 

To reflect the topological difference between domains AC and BC, we introduce a trijunction structure by connecting two domains [see inset of Fig.~\ref{fig2}(e)]. 
The energy spectrum of the trijunction structure is plotted, with the color of each eigenenergy point representing the inverse participation ratio (IPR) of the corresponding eigenstate $\psi$, defined as $\text{IPR} = \frac{\Sigma_i |\psi |^4}{(\Sigma_i |\psi |^2)^2}$. 
In Fig.~\ref{fig2}(e), we can observe a zero-energy state with the highest IPR in the bulk gap, indicating a strong localization at the trijunction, as shown in Fig.~\ref{fig2}(f). Its robustness is verified by introducing random disorders~\cite{Note1}.   
Although an additional interface exists between insulators A and B while constructing the trijunction in Fig.~\ref{fig2}(f), the projective mirror symmetry $\mathcal{M}_x$ is sufficient to protect the topological state~\cite{Note1}. Other in-gap states in Fig.~\ref{fig2}(e) with lower IPRs are trivial edge states. 
Note that the topological zero-energy state in Fig.~\ref{fig2}(f) does not arise from the higher-order topology, because of the absence of the quantized quadrupole moment of insulators A, B, and C~\cite{Benalcazar2017, Note1}. 
In other words, structures with corners formed by any pairs of insulators A, B, and C cannot induce a zero-energy state~\cite{Note1}. 

\begin{figure}[tp!]
  \centering
  \includegraphics[width=\linewidth]{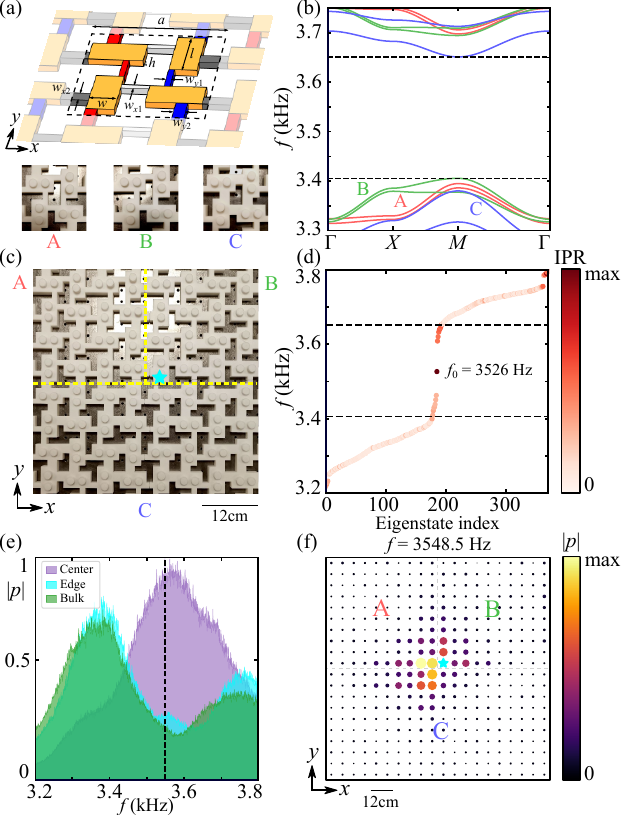}
  \caption{\textbf{Observations of the embedded topology in 2D acoustic lattices}. (a) The acoustic lattice consisting of acoustic resonators (orange) and coupling tubes. The dotted box encloses the unit cell with side length $a=120$~mm. 
(Insets) Top views of the three insulators with different tube widths~\cite{Note1}. (b) Bulk dispersions of the three insulators.
  (c) Snapshot of the finite sample with a trijunction at the center.
  (d) Calculated eigenfrequencies of the finite sample, color-coded by the IPR of each state. The dashed lines enclose the simulated common bulk bandgap.
  (e) Measured acoustic pressure as the speaker and microphone are placed at the respective regions in the sample, i.e., trijunction center, edge and bulk. The dashed line denotes the frequency of the topological state (i.e., 3526 Hz). (f) Measured distribution of the trijunction-induced topological state. The cyan star and dashed lines in (c) and (f) indicate the position of speaker and domain walls. 
  }
  \label{fig3}
\end{figure}

\begin{figure*}[tp!]
  \centering
  \includegraphics[width=\linewidth]{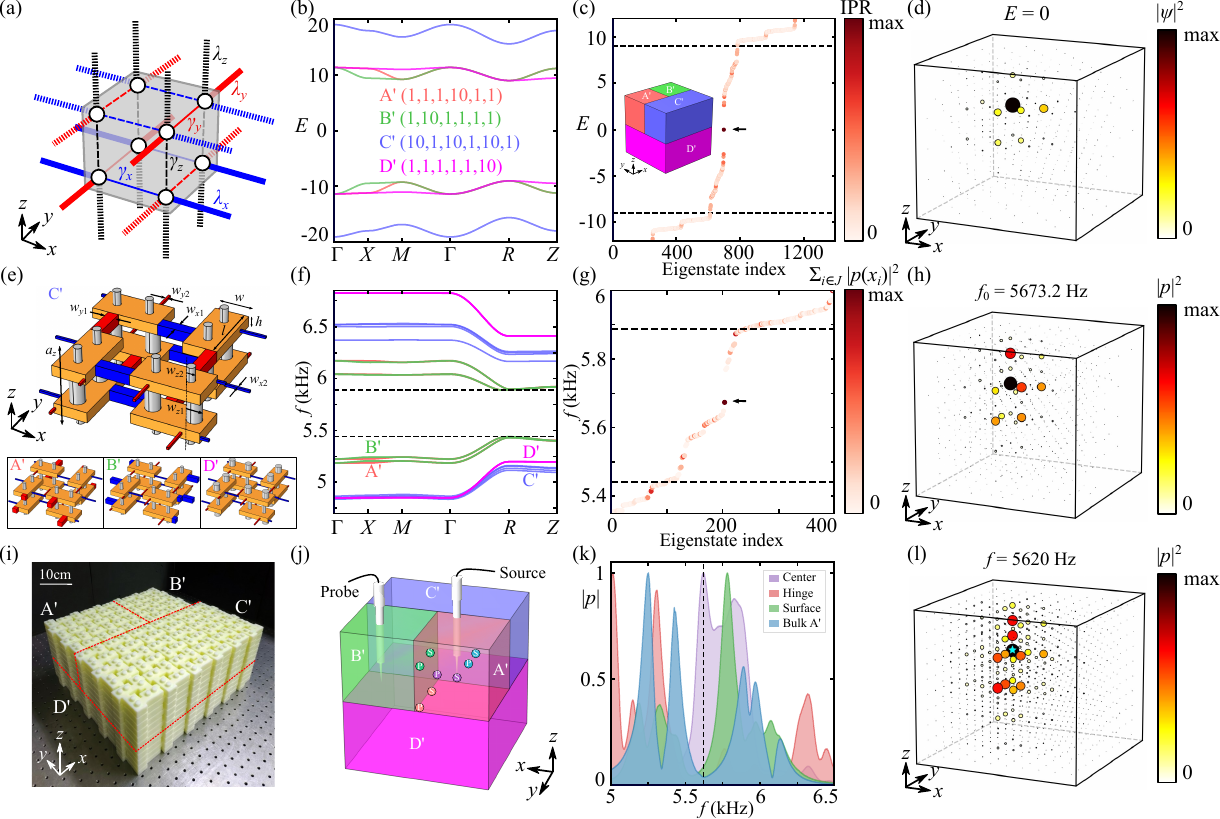}
  \caption{\textbf{Observations of the embedded topology in 3D acoustic lattices}. (a) Schematic diagram of the tight-binding model. The solid (dotted) lines indicate positive (negative) couplings, with their color and thickness representing their magnitudes.
  (b) Bulk dispersions of insulators A', B', C', and D' with different couplings $(\gamma_x, \lambda_x,\gamma_y, \lambda_y,\gamma_z, \lambda_z)$. (c) Eigenenergy spectra of the finite 3D lattice with a quadrijunction at the center (see inset). The dashed lines enclose the common bulk bandgap of the four insulators, color-coded by the IPRs of each eigenstate. (d) The topologically protected zero-energy state localized at the junction.
  (e) Unit cell of acoustic lattice, comprising 8 acoustic resonators connected by coupling tubes. Four insulators with different tube widths are shown, together with bulk dispersions in~(f).
  (g)~Calculated eigenfrequencies of the acoustic lattice with a quadrijunction formed by the same domain connection as (c). The color denotes the total intensity at 8 unit cells  (indexed by $J$) at the quadrijunction center. (h) The pressure amplitude of the topological state, with the frequency denoted by an arrow in (g). 
  (i) Snapshot of the fabricated sample. (j) Placements of sources (S) and probes (P) at different regions to obtain their respective transmission spectra as shown in (k). The transmission curves share the same colors as source-probe pairs. (l) Excitation of the topological state at the transmission peak at the quadrijunction, where the cyan star is the source position. 
  }
  \label{fig4}
\end{figure*}

Next, we experimentally observe the embedded topology in an acoustic lattice, as depicted in Fig.~\ref{fig3}(a). 
The acoustic lattice comprises of cuboid acoustic resonators (colored in orange) and coupling tubes. 
Each resonator supports a dipolar mode around 3573~Hz, whereas the dimerized couplings between resonators are realized by tuning the widths of tubes with rectangular cross sections~\cite{Xue2022,Li2022}. 
Geometric details can be found in the Supplementary Material~\cite{Note1}.
By matching the tube widths with tight-binding couplings, a common bulk bandgap is centered at 3528~Hz, with 245~Hz in size [see Fig.~\ref{fig3}(b)]. By connecting the trivial insulators, a pair of gapped interface bands emerge in domains AC and BC~\cite{Note1}. 
To demonstrate the topological properties of gapped interface bands, we construct the trijunction structure as shown in Fig.~\ref{fig3}(c), using a $20\times20$ resonator lattice with the trijunction at its center. The simulated eigenfrequencies of the finite 2D lattice are plotted in Fig.~\ref{fig3}(d). In the bandgap, an eigenstate at 3526 Hz emerges with a maximum IPR, indicating the topological distinction between domains AC and BC.

In experiments, the samples are fabricated through 3D printing according to parameters in Fig.~\ref{fig3}(a).
We measure the bulk and edge transmission spectra by placing the speaker and microphone at the respective regions of the sample~\cite{Note1}.
As depicted in Fig.~\ref{fig3}(e), both transmission spectra for edge and bulk states display two peaks, denoting a gap between 3.4~kHz and 3.7~kHz, which agrees with the simulation results in Fig.~\ref{fig3}(d).
Around the center of the trijunction, a topological state can exist in the bandgap. 
Finally, we verify the existence of the topological state by measuring the pressure distribution upon excitation at the frequency 3548.5~Hz  (approximately a $0.6\%$ deviation from the theoretical prediction) [see Fig.~\ref{fig3}(f)]. 

Furthermore, we discuss the embedded topology in 3D trivial insulators under projective crystal symmetry. 
Here, we consider the tight-binding model in Fig.~\ref{fig4}(a), which has the Hamiltonian as: $H_\text{3D} =  s_3 H_\text{2D}+
     (\gamma_z + \lambda_z \cos{k_z})\,\Gamma_5' + \lambda_z \sin{k_z} \,\Gamma_6'$, where $\Gamma_i' = s_3\Gamma_i$ ($i=1,2,3,4$), $\Gamma_5' = -s_1 \mathbb{I}_{4}$, $\Gamma_6' = -s_2\mathbb{I}_{4}$, $\{\Gamma_\mu', \Gamma_\nu'\} = 2\delta_{\mu\nu}\mathbb{I}_8$, $\gamma_z,\lambda_z>0$, and $s_i$ is the Pauli matrix. 
Similarly, we preserve the projective mirror symmetry $\mathcal{M}_x'=s_0\mathcal{M}_x$ and time-reversal symmetry $\mathcal{T}$. 
There is a common bulk gap for the four trivial 3D insulators A', B', C', and D'~\cite{Chiu2016, Chiu2013, Note1}, as depicted in Fig.~\ref{fig4}(b). 
By consecutively applying bulk-boundary correspondence on trivial bulk and lower-dimensional interfaces
(i.e., from bulk to surface, and from surface to hinge), the resulting 1D hinge Hamiltonian can exhibit topological features~\cite{Note1}, like the 2D case in Fig.~\ref{fig2}(c).
As shown in the inset of Fig.~\ref{fig4}(c), connecting the four insulators  leads to the emergence of a topological zero-energy state in the bulk gap with maximum IPR. 
This state is localized at the quadrijunction, which is at the inner vertex shared by all four insulators [see Fig.~\ref{fig4}(d)].
Note that although the model in Fig.~\ref{fig4}(a) is similar to the octupole TI \cite{Benalcazar2017, Xue2020, Ni2020}, none of the four insulators is in the octupole TI phase. 

We design an acoustic lattice as illustrated in Fig.~\ref{fig4}(e). According to Ref.~\cite{Xue2020}, the lattice consists of acoustic resonators supporting dipolar mode around 5633~Hz. The resonators are connected by thin tubes, where similar choices of thickness as Fig.~\ref{fig4}(b) lead to a common bulk bandgap from 5440~Hz to 5888~Hz [see Fig.~\ref{fig4}(f)]. 
By performing insulator connections in Fig.~\ref{fig4}(c), an in-gap topological state emerges at 5673.2~Hz [see Fig.~\ref{fig4}(g)]. 
As shown in Fig.~\ref{fig4}(h), this state is localized at the quadrijunction. 
To experimentally verify the quadrijunction-induced topological state, we fabricate the sample with similar insulator connections via 3D printing [see Fig.~\ref{fig4}(i)]. The source and probe are placed in the same region (i.e., bulk, surface, hinge, or center) as labeled in Fig.~\ref{fig4}(j). The resultant transmission spectra are plotted in Fig.~\ref{fig4}(k). 
As shown in Fig.~\ref{fig4}(k), there is a state with a pronounced peak at 5620~Hz in the common bulk gap. 
From the pressure field distribution in Fig.~\ref{fig4}(l), we can see that this state is strongly localized at the quadrijunction center, which agrees with the theoretical predictions.

Finally, we briefly differentiate between the embedded topology and topological defects. 
For example, the 2D Dirac vortex~\cite{Hou2007, Iadecola2016, Gao2019, Chen2019, Noh2020, Menssen2020, Gao2020, Lin2020} (or, 3D monopole modes~\cite{Cheng2024}) is protected by chiral symmetry. 
Our work relies on projective crystal symmetry whose $M\mathbb{Z}$ topology arises from mirror invariant planes, which goes beyond the Altland-Zirnbauer classification for topological defects~\cite{teo2010topological}.  
Furthermore, our work constructs topological states via dimension reduction, whereas the Dirac vortex usually requires a singularity in continuous parameter modulation.
More details can be found in Supplementary Material~\cite{Note1}.

\begin{acknowledgments}
To summarize, we have demonstrated embedded topology in a trivial bulk enabled by projective crystal symmetry, thereby experimentally establishing an unconventional bulk-boundary correspondence that can be initiated from a topologically trivial bulk.
Our approach can be readily extended to other wave platforms~\cite{SerraGarcia2018, Keil2016, Peterson2018, Mittal2019, Qi2020, He2020} and electric circuits~\cite{Imhof2018, Lee2018, Bao2019, Liu2020, Yamada2022}. Moreover, the interplay of non-Hermiticity-induced phase transitions~\cite{Liu2019,Lee2019, Luo2019, Kawabata2019,Kawabata2019a, Gao2021} and nonlinearity-controlled couplings~\cite{SerraGarcia2019,ZangenehNejad2019,Wang2019a,Kirsch2021} offers further opportunities for constructing embedded topological phases.
Our work also provides a fresh perspective on topological triviality and unlocks additional degrees of freedom for the design of robust topological devices~\cite{Lu2018, Queiroz2019, Noh2020, Menssen2020, Gao2020,Lin2020, Ota2018, Kim2020, Smirnova2020,Laforge2021, Lin2023, Ma2024}.

H.T.T., Y.L., K.S., and B.Z. are supported by Singapore National Research Foundation Competitive Research Program under Grant No. NRF-CRP23-2019-0007, Singapore Ministry of Education Academic Research Fund Tier 2 under Grant No. MOE-T2EP50123-0007, and Tier 1 under Grant No. RG139/22 and RG81/23. 
H.-x.S. and Y.G. are supported by the National Natural Science Foundation of China (Grants Nos. 12274183 and 12174159). S.-q.Y. is supported by the National Key R\&D Program of China (Grant No. 2020YFC1512403).
H.X. acknowledges support from the start-up fund and the direct grant 4053675 of The Chinese University of Hong Kong.
Y.L. gratefully acknowledges the support of the Eric and Wendy Schmidt AI in Science Postdoctoral Fellowship, a Schmidt Futures program. 
\end{acknowledgments}


\begin{thebibliography}{70}%
\makeatletter
\providecommand \@ifxundefined [1]{%
 \@ifx{#1\undefined}
}%
\providecommand \@ifnum [1]{%
 \ifnum #1\expandafter \@firstoftwo
 \else \expandafter \@secondoftwo
 \fi
}%
\providecommand \@ifx [1]{%
 \ifx #1\expandafter \@firstoftwo
 \else \expandafter \@secondoftwo
 \fi
}%
\providecommand \natexlab [1]{#1}%
\providecommand \enquote  [1]{``#1''}%
\providecommand \bibnamefont  [1]{#1}%
\providecommand \bibfnamefont [1]{#1}%
\providecommand \citenamefont [1]{#1}%
\providecommand \href@noop [0]{\@secondoftwo}%
\providecommand \href [0]{\begingroup \@sanitize@url \@href}%
\providecommand \@href[1]{\@@startlink{#1}\@@href}%
\providecommand \@@href[1]{\endgroup#1\@@endlink}%
\providecommand \@sanitize@url [0]{\catcode `\\12\catcode `\$12\catcode
  `\&12\catcode `\#12\catcode `\^12\catcode `\_12\catcode `\%12\relax}%
\providecommand \@@startlink[1]{}%
\providecommand \@@endlink[0]{}%
\providecommand \url  [0]{\begingroup\@sanitize@url \@url }%
\providecommand \@url [1]{\endgroup\@href {#1}{\urlprefix }}%
\providecommand \urlprefix  [0]{URL }%
\providecommand \Eprint [0]{\href }%
\providecommand \doibase [0]{https://doi.org/}%
\providecommand \selectlanguage [0]{\@gobble}%
\providecommand \bibinfo  [0]{\@secondoftwo}%
\providecommand \bibfield  [0]{\@secondoftwo}%
\providecommand \translation [1]{[#1]}%
\providecommand \BibitemOpen [0]{}%
\providecommand \bibitemStop [0]{}%
\providecommand \bibitemNoStop [0]{.\EOS\space}%
\providecommand \EOS [0]{\spacefactor3000\relax}%
\providecommand \BibitemShut  [1]{\csname bibitem#1\endcsname}%
\let\auto@bib@innerbib\@empty
\bibitem [{\citenamefont {Klitzing}\ \emph {et~al.}(1980)\citenamefont
  {Klitzing}, \citenamefont {Dorda},\ and\ \citenamefont
  {Pepper}}]{Klitzing1980}%
  \BibitemOpen
  \bibfield  {author} {\bibinfo {author} {\bibfnamefont {K.~v.}\ \bibnamefont
  {Klitzing}}, \bibinfo {author} {\bibfnamefont {G.}~\bibnamefont {Dorda}},\
  and\ \bibinfo {author} {\bibfnamefont {M.}~\bibnamefont {Pepper}},\
  }\bibfield  {title} {\bibinfo {title} {New method for high-accuracy
  determination of the fine-structure constant based on quantized {H}all
  resistance},\ }\href {https://doi.org/10.1103/physrevlett.45.494} {\bibfield
  {journal} {\bibinfo  {journal} {Phys. Rev. Lett.}\ }\textbf {\bibinfo
  {volume} {45}},\ \bibinfo {pages} {494} (\bibinfo {year} {1980})}\BibitemShut
  {NoStop}%
\bibitem [{\citenamefont {Halperin}(1982)}]{Halperin1982}%
  \BibitemOpen
  \bibfield  {author} {\bibinfo {author} {\bibfnamefont {B.~I.}\ \bibnamefont
  {Halperin}},\ }\bibfield  {title} {\bibinfo {title} {Quantized {H}all
  conductance, current-carrying edge states, and the existence of extended
  states in a two-dimensional disordered potential},\ }\href
  {https://doi.org/10.1103/physrevb.25.2185} {\bibfield  {journal} {\bibinfo
  {journal} {Phys. Rev. B}\ }\textbf {\bibinfo {volume} {25}},\ \bibinfo
  {pages} {2185} (\bibinfo {year} {1982})}\BibitemShut {NoStop}%
\bibitem [{\citenamefont {Hasan}\ and\ \citenamefont {Kane}(2010)}]{Hasan2010}%
  \BibitemOpen
  \bibfield  {author} {\bibinfo {author} {\bibfnamefont {M.~Z.}\ \bibnamefont
  {Hasan}}\ and\ \bibinfo {author} {\bibfnamefont {C.~L.}\ \bibnamefont
  {Kane}},\ }\bibfield  {title} {\bibinfo {title} {Colloquium: Topological
  insulators},\ }\href {https://doi.org/10.1103/revmodphys.82.3045} {\bibfield
  {journal} {\bibinfo  {journal} {Rev. Mod. Phys.}\ }\textbf {\bibinfo {volume}
  {82}},\ \bibinfo {pages} {3045} (\bibinfo {year} {2010})}\BibitemShut
  {NoStop}%
\bibitem [{\citenamefont {Bansil}\ \emph {et~al.}(2016)\citenamefont {Bansil},
  \citenamefont {Lin},\ and\ \citenamefont {Das}}]{Bansil2016}%
  \BibitemOpen
  \bibfield  {author} {\bibinfo {author} {\bibfnamefont {A.}~\bibnamefont
  {Bansil}}, \bibinfo {author} {\bibfnamefont {H.}~\bibnamefont {Lin}},\ and\
  \bibinfo {author} {\bibfnamefont {T.}~\bibnamefont {Das}},\ }\bibfield
  {title} {\bibinfo {title} {Colloquium: Topological band theory},\ }\href
  {https://doi.org/10.1103/revmodphys.88.021004} {\bibfield  {journal}
  {\bibinfo  {journal} {Rev. Mod. Phys.}\ }\textbf {\bibinfo {volume} {88}},\
  \bibinfo {pages} {021004} (\bibinfo {year} {2016})}\BibitemShut {NoStop}%
\bibitem [{\citenamefont {Chiu}\ \emph {et~al.}(2016)\citenamefont {Chiu},
  \citenamefont {Teo}, \citenamefont {Schnyder},\ and\ \citenamefont
  {Ryu}}]{Chiu2016}%
  \BibitemOpen
  \bibfield  {author} {\bibinfo {author} {\bibfnamefont {C.-K.}\ \bibnamefont
  {Chiu}}, \bibinfo {author} {\bibfnamefont {J.~C.~Y.}\ \bibnamefont {Teo}},
  \bibinfo {author} {\bibfnamefont {A.~P.}\ \bibnamefont {Schnyder}},\ and\
  \bibinfo {author} {\bibfnamefont {S.}~\bibnamefont {Ryu}},\ }\bibfield
  {title} {\bibinfo {title} {Classification of topological quantum matter with
  symmetries},\ }\href {https://doi.org/10.1103/revmodphys.88.035005}
  {\bibfield  {journal} {\bibinfo  {journal} {Rev. Mod. Phys.}\ }\textbf
  {\bibinfo {volume} {88}},\ \bibinfo {pages} {035005} (\bibinfo {year}
  {2016})}\BibitemShut {NoStop}%
\bibitem [{\citenamefont {Wang}\ \emph {et~al.}(2009)\citenamefont {Wang},
  \citenamefont {Chong}, \citenamefont {Joannopoulos},\ and\ \citenamefont
  {Soljačić}}]{Wang2009}%
  \BibitemOpen
  \bibfield  {author} {\bibinfo {author} {\bibfnamefont {Z.}~\bibnamefont
  {Wang}}, \bibinfo {author} {\bibfnamefont {Y.}~\bibnamefont {Chong}},
  \bibinfo {author} {\bibfnamefont {J.~D.}\ \bibnamefont {Joannopoulos}},\ and\
  \bibinfo {author} {\bibfnamefont {M.}~\bibnamefont {Soljačić}},\ }\bibfield
   {title} {\bibinfo {title} {Observation of unidirectional
  backscattering-immune topological electromagnetic states},\ }\href
  {https://doi.org/10.1038/nature08293} {\bibfield  {journal} {\bibinfo
  {journal} {Nature (London)}\ }\textbf {\bibinfo {volume} {461}},\ \bibinfo
  {pages} {772} (\bibinfo {year} {2009})}\BibitemShut {NoStop}%
\bibitem [{\citenamefont {Lu}\ \emph {et~al.}(2014)\citenamefont {Lu},
  \citenamefont {Joannopoulos},\ and\ \citenamefont {Soljačić}}]{Lu2014}%
  \BibitemOpen
  \bibfield  {author} {\bibinfo {author} {\bibfnamefont {L.}~\bibnamefont
  {Lu}}, \bibinfo {author} {\bibfnamefont {J.~D.}\ \bibnamefont
  {Joannopoulos}},\ and\ \bibinfo {author} {\bibfnamefont {M.}~\bibnamefont
  {Soljačić}},\ }\bibfield  {title} {\bibinfo {title} {Topological
  photonics},\ }\href {https://doi.org/10.1038/nphoton.2014.248} {\bibfield
  {journal} {\bibinfo  {journal} {Nat. Photonics}\ }\textbf {\bibinfo {volume}
  {8}},\ \bibinfo {pages} {821} (\bibinfo {year} {2014})}\BibitemShut {NoStop}%
\bibitem [{\citenamefont {Yang}\ \emph {et~al.}(2015)\citenamefont {Yang},
  \citenamefont {Gao}, \citenamefont {Shi}, \citenamefont {Lin}, \citenamefont
  {Gao}, \citenamefont {Chong},\ and\ \citenamefont {Zhang}}]{Yang2015}%
  \BibitemOpen
  \bibfield  {author} {\bibinfo {author} {\bibfnamefont {Z.}~\bibnamefont
  {Yang}}, \bibinfo {author} {\bibfnamefont {F.}~\bibnamefont {Gao}}, \bibinfo
  {author} {\bibfnamefont {X.}~\bibnamefont {Shi}}, \bibinfo {author}
  {\bibfnamefont {X.}~\bibnamefont {Lin}}, \bibinfo {author} {\bibfnamefont
  {Z.}~\bibnamefont {Gao}}, \bibinfo {author} {\bibfnamefont {Y.}~\bibnamefont
  {Chong}},\ and\ \bibinfo {author} {\bibfnamefont {B.}~\bibnamefont {Zhang}},\
  }\bibfield  {title} {\bibinfo {title} {Topological acoustics},\ }\href
  {https://doi.org/10.1103/physrevlett.114.114301} {\bibfield  {journal}
  {\bibinfo  {journal} {Phys. Rev. Lett.}\ }\textbf {\bibinfo {volume} {114}},\
  \bibinfo {pages} {114301} (\bibinfo {year} {2015})}\BibitemShut {NoStop}%
\bibitem [{\citenamefont {Benalcazar}\ \emph
  {et~al.}(2017{\natexlab{a}})\citenamefont {Benalcazar}, \citenamefont
  {Bernevig},\ and\ \citenamefont {Hughes}}]{Benalcazar2017}%
  \BibitemOpen
  \bibfield  {author} {\bibinfo {author} {\bibfnamefont {W.~A.}\ \bibnamefont
  {Benalcazar}}, \bibinfo {author} {\bibfnamefont {B.~A.}\ \bibnamefont
  {Bernevig}},\ and\ \bibinfo {author} {\bibfnamefont {T.~L.}\ \bibnamefont
  {Hughes}},\ }\bibfield  {title} {\bibinfo {title} {Quantized electric
  multipole insulators},\ }\href {https://doi.org/10.1126/science.aah6442}
  {\bibfield  {journal} {\bibinfo  {journal} {Science}\ }\textbf {\bibinfo
  {volume} {357}},\ \bibinfo {pages} {61} (\bibinfo {year}
  {2017}{\natexlab{a}})}\BibitemShut {NoStop}%
\bibitem [{\citenamefont {Benalcazar}\ \emph
  {et~al.}(2017{\natexlab{b}})\citenamefont {Benalcazar}, \citenamefont
  {Bernevig},\ and\ \citenamefont {Hughes}}]{Benalcazar2017a}%
  \BibitemOpen
  \bibfield  {author} {\bibinfo {author} {\bibfnamefont {W.~A.}\ \bibnamefont
  {Benalcazar}}, \bibinfo {author} {\bibfnamefont {B.~A.}\ \bibnamefont
  {Bernevig}},\ and\ \bibinfo {author} {\bibfnamefont {T.~L.}\ \bibnamefont
  {Hughes}},\ }\bibfield  {title} {\bibinfo {title} {Electric multipole
  moments, topological multipole moment pumping, and chiral hinge states in
  crystalline insulators},\ }\href {https://doi.org/10.1103/physrevb.96.245115}
  {\bibfield  {journal} {\bibinfo  {journal} {Phys. Rev. B}\ }\textbf {\bibinfo
  {volume} {96}},\ \bibinfo {pages} {245115} (\bibinfo {year}
  {2017}{\natexlab{b}})}\BibitemShut {NoStop}%
\bibitem [{\citenamefont {Serra-Garcia}\ \emph {et~al.}(2018)\citenamefont
  {Serra-Garcia}, \citenamefont {Peri}, \citenamefont {Süsstrunk},
  \citenamefont {Bilal}, \citenamefont {Larsen}, \citenamefont {Villanueva},\
  and\ \citenamefont {Huber}}]{SerraGarcia2018}%
  \BibitemOpen
  \bibfield  {author} {\bibinfo {author} {\bibfnamefont {M.}~\bibnamefont
  {Serra-Garcia}}, \bibinfo {author} {\bibfnamefont {V.}~\bibnamefont {Peri}},
  \bibinfo {author} {\bibfnamefont {R.}~\bibnamefont {Süsstrunk}}, \bibinfo
  {author} {\bibfnamefont {O.~R.}\ \bibnamefont {Bilal}}, \bibinfo {author}
  {\bibfnamefont {T.}~\bibnamefont {Larsen}}, \bibinfo {author} {\bibfnamefont
  {L.~G.}\ \bibnamefont {Villanueva}},\ and\ \bibinfo {author} {\bibfnamefont
  {S.~D.}\ \bibnamefont {Huber}},\ }\bibfield  {title} {\bibinfo {title}
  {Observation of a phononic quadrupole topological insulator},\ }\href
  {https://doi.org/10.1038/nature25156} {\bibfield  {journal} {\bibinfo
  {journal} {Nature (London)}\ }\textbf {\bibinfo {volume} {555}},\ \bibinfo
  {pages} {342} (\bibinfo {year} {2018})}\BibitemShut {NoStop}%
\bibitem [{\citenamefont {Peterson}\ \emph {et~al.}(2018)\citenamefont
  {Peterson}, \citenamefont {Benalcazar}, \citenamefont {Hughes},\ and\
  \citenamefont {Bahl}}]{Peterson2018}%
  \BibitemOpen
  \bibfield  {author} {\bibinfo {author} {\bibfnamefont {C.~W.}\ \bibnamefont
  {Peterson}}, \bibinfo {author} {\bibfnamefont {W.~A.}\ \bibnamefont
  {Benalcazar}}, \bibinfo {author} {\bibfnamefont {T.~L.}\ \bibnamefont
  {Hughes}},\ and\ \bibinfo {author} {\bibfnamefont {G.}~\bibnamefont {Bahl}},\
  }\bibfield  {title} {\bibinfo {title} {A quantized microwave quadrupole
  insulator with topologically protected corner states},\ }\href
  {https://doi.org/10.1038/nature25777} {\bibfield  {journal} {\bibinfo
  {journal} {Nature (London)}\ }\textbf {\bibinfo {volume} {555}},\ \bibinfo
  {pages} {346} (\bibinfo {year} {2018})}\BibitemShut {NoStop}%
\bibitem [{\citenamefont {Mittal}\ \emph {et~al.}(2019)\citenamefont {Mittal},
  \citenamefont {Orre}, \citenamefont {Zhu}, \citenamefont {Gorlach},
  \citenamefont {Poddubny},\ and\ \citenamefont {Hafezi}}]{Mittal2019}%
  \BibitemOpen
  \bibfield  {author} {\bibinfo {author} {\bibfnamefont {S.}~\bibnamefont
  {Mittal}}, \bibinfo {author} {\bibfnamefont {V.~V.}\ \bibnamefont {Orre}},
  \bibinfo {author} {\bibfnamefont {G.}~\bibnamefont {Zhu}}, \bibinfo {author}
  {\bibfnamefont {M.~A.}\ \bibnamefont {Gorlach}}, \bibinfo {author}
  {\bibfnamefont {A.}~\bibnamefont {Poddubny}},\ and\ \bibinfo {author}
  {\bibfnamefont {M.}~\bibnamefont {Hafezi}},\ }\bibfield  {title} {\bibinfo
  {title} {Photonic quadrupole topological phases},\ }\href
  {https://doi.org/10.1038/s41566-019-0452-0} {\bibfield  {journal} {\bibinfo
  {journal} {Nat. Photonics}\ }\textbf {\bibinfo {volume} {13}},\ \bibinfo
  {pages} {692} (\bibinfo {year} {2019})}\BibitemShut {NoStop}%
\bibitem [{\citenamefont {Qi}\ \emph {et~al.}(2020)\citenamefont {Qi},
  \citenamefont {Qiu}, \citenamefont {Xiao}, \citenamefont {He}, \citenamefont
  {Ke},\ and\ \citenamefont {Liu}}]{Qi2020}%
  \BibitemOpen
  \bibfield  {author} {\bibinfo {author} {\bibfnamefont {Y.}~\bibnamefont
  {Qi}}, \bibinfo {author} {\bibfnamefont {C.}~\bibnamefont {Qiu}}, \bibinfo
  {author} {\bibfnamefont {M.}~\bibnamefont {Xiao}}, \bibinfo {author}
  {\bibfnamefont {H.}~\bibnamefont {He}}, \bibinfo {author} {\bibfnamefont
  {M.}~\bibnamefont {Ke}},\ and\ \bibinfo {author} {\bibfnamefont
  {Z.}~\bibnamefont {Liu}},\ }\bibfield  {title} {\bibinfo {title} {Acoustic
  realization of quadrupole topological insulators},\ }\href
  {https://doi.org/10.1103/physrevlett.124.206601} {\bibfield  {journal}
  {\bibinfo  {journal} {Phys. Rev. Lett.}\ }\textbf {\bibinfo {volume} {124}},\
  \bibinfo {pages} {206601} (\bibinfo {year} {2020})}\BibitemShut {NoStop}%
\bibitem [{\citenamefont {He}\ \emph {et~al.}(2020)\citenamefont {He},
  \citenamefont {Addison}, \citenamefont {Mele},\ and\ \citenamefont
  {Zhen}}]{He2020}%
  \BibitemOpen
  \bibfield  {author} {\bibinfo {author} {\bibfnamefont {L.}~\bibnamefont
  {He}}, \bibinfo {author} {\bibfnamefont {Z.}~\bibnamefont {Addison}},
  \bibinfo {author} {\bibfnamefont {E.~J.}\ \bibnamefont {Mele}},\ and\
  \bibinfo {author} {\bibfnamefont {B.}~\bibnamefont {Zhen}},\ }\bibfield
  {title} {\bibinfo {title} {Quadrupole topological photonic crystals},\ }\href
  {https://doi.org/10.1038/s41467-020-16916-z} {\bibfield  {journal} {\bibinfo
  {journal} {Nat. Commun.}\ }\textbf {\bibinfo {volume} {11}},\ \bibinfo
  {pages} {3119} (\bibinfo {year} {2020})}\BibitemShut {NoStop}%
\bibitem [{\citenamefont {Xue}\ \emph {et~al.}(2020)\citenamefont {Xue},
  \citenamefont {Ge}, \citenamefont {Sun}, \citenamefont {Wang}, \citenamefont
  {Jia}, \citenamefont {Guan}, \citenamefont {Yuan}, \citenamefont {Chong},\
  and\ \citenamefont {Zhang}}]{Xue2020}%
  \BibitemOpen
  \bibfield  {author} {\bibinfo {author} {\bibfnamefont {H.}~\bibnamefont
  {Xue}}, \bibinfo {author} {\bibfnamefont {Y.}~\bibnamefont {Ge}}, \bibinfo
  {author} {\bibfnamefont {H.-X.}\ \bibnamefont {Sun}}, \bibinfo {author}
  {\bibfnamefont {Q.}~\bibnamefont {Wang}}, \bibinfo {author} {\bibfnamefont
  {D.}~\bibnamefont {Jia}}, \bibinfo {author} {\bibfnamefont {Y.-J.}\
  \bibnamefont {Guan}}, \bibinfo {author} {\bibfnamefont {S.-Q.}\ \bibnamefont
  {Yuan}}, \bibinfo {author} {\bibfnamefont {Y.}~\bibnamefont {Chong}},\ and\
  \bibinfo {author} {\bibfnamefont {B.}~\bibnamefont {Zhang}},\ }\bibfield
  {title} {\bibinfo {title} {Observation of an acoustic octupole topological
  insulator},\ }\href {https://doi.org/10.1038/s41467-020-16350-1} {\bibfield
  {journal} {\bibinfo  {journal} {Nat. Commun.}\ }\textbf {\bibinfo {volume}
  {11}},\ \bibinfo {pages} {2442} (\bibinfo {year} {2020})}\BibitemShut
  {NoStop}%
\bibitem [{\citenamefont {Ni}\ \emph {et~al.}(2020)\citenamefont {Ni},
  \citenamefont {Li}, \citenamefont {Weiner}, \citenamefont {Alù},\ and\
  \citenamefont {Khanikaev}}]{Ni2020}%
  \BibitemOpen
  \bibfield  {author} {\bibinfo {author} {\bibfnamefont {X.}~\bibnamefont
  {Ni}}, \bibinfo {author} {\bibfnamefont {M.}~\bibnamefont {Li}}, \bibinfo
  {author} {\bibfnamefont {M.}~\bibnamefont {Weiner}}, \bibinfo {author}
  {\bibfnamefont {A.}~\bibnamefont {Alù}},\ and\ \bibinfo {author}
  {\bibfnamefont {A.~B.}\ \bibnamefont {Khanikaev}},\ }\bibfield  {title}
  {\bibinfo {title} {Demonstration of a quantized acoustic octupole topological
  insulator},\ }\href {https://doi.org/10.1038/s41467-020-15705-y} {\bibfield
  {journal} {\bibinfo  {journal} {Nat. Commun.}\ }\textbf {\bibinfo {volume}
  {11}},\ \bibinfo {pages} {2108} (\bibinfo {year} {2020})}\BibitemShut
  {NoStop}%
\bibitem [{\citenamefont {Xie}\ \emph {et~al.}(2021)\citenamefont {Xie},
  \citenamefont {Wang}, \citenamefont {Zhang}, \citenamefont {Zhan},
  \citenamefont {Jiang}, \citenamefont {Lu},\ and\ \citenamefont
  {Chen}}]{Xie2021}%
  \BibitemOpen
  \bibfield  {author} {\bibinfo {author} {\bibfnamefont {B.}~\bibnamefont
  {Xie}}, \bibinfo {author} {\bibfnamefont {H.-X.}\ \bibnamefont {Wang}},
  \bibinfo {author} {\bibfnamefont {X.}~\bibnamefont {Zhang}}, \bibinfo
  {author} {\bibfnamefont {P.}~\bibnamefont {Zhan}}, \bibinfo {author}
  {\bibfnamefont {J.-H.}\ \bibnamefont {Jiang}}, \bibinfo {author}
  {\bibfnamefont {M.}~\bibnamefont {Lu}},\ and\ \bibinfo {author}
  {\bibfnamefont {Y.}~\bibnamefont {Chen}},\ }\bibfield  {title} {\bibinfo
  {title} {Higher-order band topology},\ }\href
  {https://doi.org/10.1038/s42254-021-00323-4} {\bibfield  {journal} {\bibinfo
  {journal} {Nat. Rev. Phys.}\ }\textbf {\bibinfo {volume} {3}},\ \bibinfo
  {pages} {520} (\bibinfo {year} {2021})}\BibitemShut {NoStop}%
\bibitem [{\citenamefont {Tuegel}\ \emph {et~al.}(2019)\citenamefont {Tuegel},
  \citenamefont {Chua},\ and\ \citenamefont {Hughes}}]{Tuegel2019}%
  \BibitemOpen
  \bibfield  {author} {\bibinfo {author} {\bibfnamefont {T.~I.}\ \bibnamefont
  {Tuegel}}, \bibinfo {author} {\bibfnamefont {V.}~\bibnamefont {Chua}},\ and\
  \bibinfo {author} {\bibfnamefont {T.~L.}\ \bibnamefont {Hughes}},\ }\bibfield
   {title} {\bibinfo {title} {Embedded topological insulators},\ }\href
  {https://doi.org/10.1103/physrevb.100.115126} {\bibfield  {journal} {\bibinfo
   {journal} {Phys. Rev. B}\ }\textbf {\bibinfo {volume} {100}},\ \bibinfo
  {pages} {115126} (\bibinfo {year} {2019})}\BibitemShut {NoStop}%
\bibitem [{\citenamefont {Velury}\ and\ \citenamefont
  {Hughes}(2022)}]{Velury2022}%
  \BibitemOpen
  \bibfield  {author} {\bibinfo {author} {\bibfnamefont {S.}~\bibnamefont
  {Velury}}\ and\ \bibinfo {author} {\bibfnamefont {T.~L.}\ \bibnamefont
  {Hughes}},\ }\bibfield  {title} {\bibinfo {title} {Embedded topological
  semimetals},\ }\href {https://doi.org/10.1103/physrevb.105.184105} {\bibfield
   {journal} {\bibinfo  {journal} {Phys. Rev. B}\ }\textbf {\bibinfo {volume}
  {105}},\ \bibinfo {pages} {184105} (\bibinfo {year} {2022})}\BibitemShut
  {NoStop}%
\bibitem [{\citenamefont {Panigrahi}\ \emph {et~al.}(2022)\citenamefont
  {Panigrahi}, \citenamefont {Juričić},\ and\ \citenamefont
  {Roy}}]{Panigrahi2022}%
  \BibitemOpen
  \bibfield  {author} {\bibinfo {author} {\bibfnamefont {A.}~\bibnamefont
  {Panigrahi}}, \bibinfo {author} {\bibfnamefont {V.}~\bibnamefont
  {Juričić}},\ and\ \bibinfo {author} {\bibfnamefont {B.}~\bibnamefont
  {Roy}},\ }\bibfield  {title} {\bibinfo {title} {Projected topological
  branes},\ }\href {https://doi.org/10.1038/s42005-022-01006-x} {\bibfield
  {journal} {\bibinfo  {journal} {Commun. Phys.}\ }\textbf {\bibinfo {volume}
  {5}},\ \bibinfo {pages} {230} (\bibinfo {year} {2022})}\BibitemShut {NoStop}%
\bibitem [{\citenamefont {Zhao}\ \emph {et~al.}(2020)\citenamefont {Zhao},
  \citenamefont {Huang},\ and\ \citenamefont {Yang}}]{Zhao2020}%
  \BibitemOpen
  \bibfield  {author} {\bibinfo {author} {\bibfnamefont {Y.~X.}\ \bibnamefont
  {Zhao}}, \bibinfo {author} {\bibfnamefont {Y.-X.}\ \bibnamefont {Huang}},\
  and\ \bibinfo {author} {\bibfnamefont {S.~A.}\ \bibnamefont {Yang}},\
  }\bibfield  {title} {\bibinfo {title} {$\mathbb{Z}_2$-projective
  translational symmetry protected topological phases},\ }\href
  {https://doi.org/10.1103/physrevb.102.161117} {\bibfield  {journal} {\bibinfo
   {journal} {Phys. Rev. B}\ }\textbf {\bibinfo {volume} {102}},\ \bibinfo
  {pages} {161117(R)} (\bibinfo {year} {2020})}\BibitemShut {NoStop}%
\bibitem [{\citenamefont {Xue}\ \emph {et~al.}(2022)\citenamefont {Xue},
  \citenamefont {Wang}, \citenamefont {Huang}, \citenamefont {Cheng},
  \citenamefont {Yu}, \citenamefont {Foo}, \citenamefont {Zhao}, \citenamefont
  {Yang},\ and\ \citenamefont {Zhang}}]{Xue2022}%
  \BibitemOpen
  \bibfield  {author} {\bibinfo {author} {\bibfnamefont {H.}~\bibnamefont
  {Xue}}, \bibinfo {author} {\bibfnamefont {Z.}~\bibnamefont {Wang}}, \bibinfo
  {author} {\bibfnamefont {Y.-X.}\ \bibnamefont {Huang}}, \bibinfo {author}
  {\bibfnamefont {Z.}~\bibnamefont {Cheng}}, \bibinfo {author} {\bibfnamefont
  {L.}~\bibnamefont {Yu}}, \bibinfo {author} {\bibfnamefont {Y.~X.}\
  \bibnamefont {Foo}}, \bibinfo {author} {\bibfnamefont {Y.~X.}\ \bibnamefont
  {Zhao}}, \bibinfo {author} {\bibfnamefont {S.~A.}\ \bibnamefont {Yang}},\
  and\ \bibinfo {author} {\bibfnamefont {B.}~\bibnamefont {Zhang}},\ }\bibfield
   {title} {\bibinfo {title} {Projectively enriched symmetry and topology in
  acoustic crystals},\ }\href {https://doi.org/10.1103/physrevlett.128.116802}
  {\bibfield  {journal} {\bibinfo  {journal} {Phys. Rev. Lett.}\ }\textbf
  {\bibinfo {volume} {128}},\ \bibinfo {pages} {116802} (\bibinfo {year}
  {2022})}\BibitemShut {NoStop}%
\bibitem [{\citenamefont {Li}\ \emph {et~al.}(2022)\citenamefont {Li},
  \citenamefont {Du}, \citenamefont {Zhang}, \citenamefont {Li}, \citenamefont
  {Fan}, \citenamefont {Zhang},\ and\ \citenamefont {Qiu}}]{Li2022}%
  \BibitemOpen
  \bibfield  {author} {\bibinfo {author} {\bibfnamefont {T.}~\bibnamefont
  {Li}}, \bibinfo {author} {\bibfnamefont {J.}~\bibnamefont {Du}}, \bibinfo
  {author} {\bibfnamefont {Q.}~\bibnamefont {Zhang}}, \bibinfo {author}
  {\bibfnamefont {Y.}~\bibnamefont {Li}}, \bibinfo {author} {\bibfnamefont
  {X.}~\bibnamefont {Fan}}, \bibinfo {author} {\bibfnamefont {F.}~\bibnamefont
  {Zhang}},\ and\ \bibinfo {author} {\bibfnamefont {C.}~\bibnamefont {Qiu}},\
  }\bibfield  {title} {\bibinfo {title} {Acoustic {M}öbius insulators from
  projective symmetry},\ }\href
  {https://doi.org/10.1103/physrevlett.128.116803} {\bibfield  {journal}
  {\bibinfo  {journal} {Phys. Rev. Lett.}\ }\textbf {\bibinfo {volume} {128}},\
  \bibinfo {pages} {116803} (\bibinfo {year} {2022})}\BibitemShut {NoStop}%
\bibitem [{\citenamefont {Shao}\ \emph {et~al.}(2021)\citenamefont {Shao},
  \citenamefont {Liu}, \citenamefont {Xiao}, \citenamefont {Yang},\ and\
  \citenamefont {Zhao}}]{Shao2021}%
  \BibitemOpen
  \bibfield  {author} {\bibinfo {author} {\bibfnamefont {L.~B.}\ \bibnamefont
  {Shao}}, \bibinfo {author} {\bibfnamefont {Q.}~\bibnamefont {Liu}}, \bibinfo
  {author} {\bibfnamefont {R.}~\bibnamefont {Xiao}}, \bibinfo {author}
  {\bibfnamefont {S.~A.}\ \bibnamefont {Yang}},\ and\ \bibinfo {author}
  {\bibfnamefont {Y.~X.}\ \bibnamefont {Zhao}},\ }\bibfield  {title} {\bibinfo
  {title} {Gauge-field extended $k\cdot p$ method and novel topological
  phases},\ }\href {https://doi.org/10.1103/physrevlett.127.076401} {\bibfield
  {journal} {\bibinfo  {journal} {Phys. Rev. Lett.}\ }\textbf {\bibinfo
  {volume} {127}},\ \bibinfo {pages} {076401} (\bibinfo {year}
  {2021})}\BibitemShut {NoStop}%
\bibitem [{\citenamefont {Xue}\ \emph {et~al.}(2023)\citenamefont {Xue},
  \citenamefont {Chen}, \citenamefont {Cheng}, \citenamefont {Dai},
  \citenamefont {Long}, \citenamefont {Zhao},\ and\ \citenamefont
  {Zhang}}]{Xue2023}%
  \BibitemOpen
  \bibfield  {author} {\bibinfo {author} {\bibfnamefont {H.}~\bibnamefont
  {Xue}}, \bibinfo {author} {\bibfnamefont {Z.~Y.}\ \bibnamefont {Chen}},
  \bibinfo {author} {\bibfnamefont {Z.}~\bibnamefont {Cheng}}, \bibinfo
  {author} {\bibfnamefont {J.~X.}\ \bibnamefont {Dai}}, \bibinfo {author}
  {\bibfnamefont {Y.}~\bibnamefont {Long}}, \bibinfo {author} {\bibfnamefont
  {Y.~X.}\ \bibnamefont {Zhao}},\ and\ \bibinfo {author} {\bibfnamefont
  {B.}~\bibnamefont {Zhang}},\ }\bibfield  {title} {\bibinfo {title}
  {Stiefel-{W}hitney topological charges in a three-dimensional acoustic
  nodal-line crystal},\ }\href {https://doi.org/10.1038/s41467-023-40252-7}
  {\bibfield  {journal} {\bibinfo  {journal} {Nat. Commun.}\ }\textbf {\bibinfo
  {volume} {14}},\ \bibinfo {pages} {4563} (\bibinfo {year}
  {2023})}\BibitemShut {NoStop}%
\bibitem [{\citenamefont {Chen}\ \emph {et~al.}(2022)\citenamefont {Chen},
  \citenamefont {Yang},\ and\ \citenamefont {Zhao}}]{Chen2022}%
  \BibitemOpen
  \bibfield  {author} {\bibinfo {author} {\bibfnamefont {Z.~Y.}\ \bibnamefont
  {Chen}}, \bibinfo {author} {\bibfnamefont {S.~A.}\ \bibnamefont {Yang}},\
  and\ \bibinfo {author} {\bibfnamefont {Y.~X.}\ \bibnamefont {Zhao}},\
  }\bibfield  {title} {\bibinfo {title} {Brillouin {K}lein bottle from
  artificial gauge fields},\ }\href
  {https://doi.org/10.1038/s41467-022-29953-7} {\bibfield  {journal} {\bibinfo
  {journal} {Nat. Commun.}\ }\textbf {\bibinfo {volume} {13}},\ \bibinfo
  {pages} {2215} (\bibinfo {year} {2022})}\BibitemShut {NoStop}%
\bibitem [{\citenamefont {Pu}\ \emph {et~al.}(2023)\citenamefont {Pu},
  \citenamefont {He}, \citenamefont {Deng}, \citenamefont {Huang},
  \citenamefont {Ye}, \citenamefont {Lu}, \citenamefont {Ke},\ and\
  \citenamefont {Liu}}]{Pu2023}%
  \BibitemOpen
  \bibfield  {author} {\bibinfo {author} {\bibfnamefont {Z.}~\bibnamefont
  {Pu}}, \bibinfo {author} {\bibfnamefont {H.}~\bibnamefont {He}}, \bibinfo
  {author} {\bibfnamefont {W.}~\bibnamefont {Deng}}, \bibinfo {author}
  {\bibfnamefont {X.}~\bibnamefont {Huang}}, \bibinfo {author} {\bibfnamefont
  {L.}~\bibnamefont {Ye}}, \bibinfo {author} {\bibfnamefont {J.}~\bibnamefont
  {Lu}}, \bibinfo {author} {\bibfnamefont {M.}~\bibnamefont {Ke}},\ and\
  \bibinfo {author} {\bibfnamefont {Z.}~\bibnamefont {Liu}},\ }\bibfield
  {title} {\bibinfo {title} {Acoustic {K}lein bottle insulator},\ }\href
  {https://doi.org/10.1103/physrevb.108.l220101} {\bibfield  {journal}
  {\bibinfo  {journal} {Phys. Rev. B}\ }\textbf {\bibinfo {volume} {108}},\
  \bibinfo {pages} {L220101} (\bibinfo {year} {2023})}\BibitemShut {NoStop}%
\bibitem [{\citenamefont {Li}\ \emph {et~al.}(2023)\citenamefont {Li},
  \citenamefont {Sun}, \citenamefont {Zhang}, \citenamefont {Guo},\ and\
  \citenamefont {Trauzettel}}]{Li2023}%
  \BibitemOpen
  \bibfield  {author} {\bibinfo {author} {\bibfnamefont {C.-A.}\ \bibnamefont
  {Li}}, \bibinfo {author} {\bibfnamefont {J.}~\bibnamefont {Sun}}, \bibinfo
  {author} {\bibfnamefont {S.-B.}\ \bibnamefont {Zhang}}, \bibinfo {author}
  {\bibfnamefont {H.}~\bibnamefont {Guo}},\ and\ \bibinfo {author}
  {\bibfnamefont {B.}~\bibnamefont {Trauzettel}},\ }\bibfield  {title}
  {\bibinfo {title} {Klein-bottle quadrupole insulators and {D}irac
  semimetals},\ }\href {https://doi.org/10.1103/physrevb.108.235412} {\bibfield
   {journal} {\bibinfo  {journal} {Phys. Rev. B}\ }\textbf {\bibinfo {volume}
  {108}},\ \bibinfo {pages} {235412} (\bibinfo {year} {2023})}\BibitemShut
  {NoStop}%
\bibitem [{\citenamefont {Tao}\ \emph {et~al.}(2024)\citenamefont {Tao},
  \citenamefont {Yan}, \citenamefont {Peng}, \citenamefont {Wei}, \citenamefont
  {Cui}, \citenamefont {Yang}, \citenamefont {Chen},\ and\ \citenamefont
  {Xu}}]{Tao2024}%
  \BibitemOpen
  \bibfield  {author} {\bibinfo {author} {\bibfnamefont {Y.-L.}\ \bibnamefont
  {Tao}}, \bibinfo {author} {\bibfnamefont {M.}~\bibnamefont {Yan}}, \bibinfo
  {author} {\bibfnamefont {M.}~\bibnamefont {Peng}}, \bibinfo {author}
  {\bibfnamefont {Q.}~\bibnamefont {Wei}}, \bibinfo {author} {\bibfnamefont
  {Z.}~\bibnamefont {Cui}}, \bibinfo {author} {\bibfnamefont {S.~A.}\
  \bibnamefont {Yang}}, \bibinfo {author} {\bibfnamefont {G.}~\bibnamefont
  {Chen}},\ and\ \bibinfo {author} {\bibfnamefont {Y.}~\bibnamefont {Xu}},\
  }\bibfield  {title} {\bibinfo {title} {Higher-order {K}lein bottle
  topological insulator in three-dimensional acoustic crystals},\ }\href
  {https://doi.org/10.1103/physrevb.109.134107} {\bibfield  {journal} {\bibinfo
   {journal} {Phys. Rev. B}\ }\textbf {\bibinfo {volume} {109}},\ \bibinfo
  {pages} {134107} (\bibinfo {year} {2024})}\BibitemShut {NoStop}%
\bibitem [{\citenamefont {Zhu}\ \emph {et~al.}(2024)\citenamefont {Zhu},
  \citenamefont {Yang}, \citenamefont {Wu}, \citenamefont {Meng}, \citenamefont
  {Xi}, \citenamefont {Yan}, \citenamefont {Chen}, \citenamefont {Lu},
  \citenamefont {Huang}, \citenamefont {Deng}, \citenamefont {Shang},
  \citenamefont {Shum}, \citenamefont {Yang}, \citenamefont {Chen},
  \citenamefont {Xiang}, \citenamefont {Liu}, \citenamefont {Liu},\ and\
  \citenamefont {Gao}}]{Zhu2024}%
  \BibitemOpen
  \bibfield  {author} {\bibinfo {author} {\bibfnamefont {Z.}~\bibnamefont
  {Zhu}}, \bibinfo {author} {\bibfnamefont {L.}~\bibnamefont {Yang}}, \bibinfo
  {author} {\bibfnamefont {J.}~\bibnamefont {Wu}}, \bibinfo {author}
  {\bibfnamefont {Y.}~\bibnamefont {Meng}}, \bibinfo {author} {\bibfnamefont
  {X.}~\bibnamefont {Xi}}, \bibinfo {author} {\bibfnamefont {B.}~\bibnamefont
  {Yan}}, \bibinfo {author} {\bibfnamefont {J.}~\bibnamefont {Chen}}, \bibinfo
  {author} {\bibfnamefont {J.}~\bibnamefont {Lu}}, \bibinfo {author}
  {\bibfnamefont {X.}~\bibnamefont {Huang}}, \bibinfo {author} {\bibfnamefont
  {W.}~\bibnamefont {Deng}}, \bibinfo {author} {\bibfnamefont {C.}~\bibnamefont
  {Shang}}, \bibinfo {author} {\bibfnamefont {P.~P.}\ \bibnamefont {Shum}},
  \bibinfo {author} {\bibfnamefont {Y.}~\bibnamefont {Yang}}, \bibinfo {author}
  {\bibfnamefont {H.}~\bibnamefont {Chen}}, \bibinfo {author} {\bibfnamefont
  {K.}~\bibnamefont {Xiang}}, \bibinfo {author} {\bibfnamefont {G.-G.}\
  \bibnamefont {Liu}}, \bibinfo {author} {\bibfnamefont {Z.}~\bibnamefont
  {Liu}},\ and\ \bibinfo {author} {\bibfnamefont {Z.}~\bibnamefont {Gao}},\
  }\bibfield  {title} {\bibinfo {title} {{B}rillouin {K}lein space and
  half-turn space in three-dimensional acoustic crystals},\ }\href
  {https://doi.org/10.1016/j.scib.2024.05.003} {\bibfield  {journal} {\bibinfo
  {journal} {Sci. Bull.}\ }\textbf {\bibinfo {volume} {69}},\ \bibinfo {pages}
  {2050} (\bibinfo {year} {2024})}\BibitemShut {NoStop}%
\bibitem [{\citenamefont {Zhao}\ \emph {et~al.}(2021)\citenamefont {Zhao},
  \citenamefont {Chen}, \citenamefont {Sheng},\ and\ \citenamefont
  {Yang}}]{Zhao2021}%
  \BibitemOpen
  \bibfield  {author} {\bibinfo {author} {\bibfnamefont {Y.~X.}\ \bibnamefont
  {Zhao}}, \bibinfo {author} {\bibfnamefont {C.}~\bibnamefont {Chen}}, \bibinfo
  {author} {\bibfnamefont {X.-L.}\ \bibnamefont {Sheng}},\ and\ \bibinfo
  {author} {\bibfnamefont {S.~A.}\ \bibnamefont {Yang}},\ }\bibfield  {title}
  {\bibinfo {title} {Switching spinless and spinful topological phases with
  projective {$PT$} symmetry},\ }\href
  {https://doi.org/10.1103/physrevlett.126.196402} {\bibfield  {journal}
  {\bibinfo  {journal} {Phys. Rev. Lett.}\ }\textbf {\bibinfo {volume} {126}},\
  \bibinfo {pages} {196402} (\bibinfo {year} {2021})}\BibitemShut {NoStop}%
\bibitem [{\citenamefont {Huang}\ \emph {et~al.}(2022)\citenamefont {Huang},
  \citenamefont {Chen}, \citenamefont {Feng}, \citenamefont {Yang},\ and\
  \citenamefont {Zhao}}]{Huang2022}%
  \BibitemOpen
  \bibfield  {author} {\bibinfo {author} {\bibfnamefont {Y.-X.}\ \bibnamefont
  {Huang}}, \bibinfo {author} {\bibfnamefont {Z.~Y.}\ \bibnamefont {Chen}},
  \bibinfo {author} {\bibfnamefont {X.}~\bibnamefont {Feng}}, \bibinfo {author}
  {\bibfnamefont {S.~A.}\ \bibnamefont {Yang}},\ and\ \bibinfo {author}
  {\bibfnamefont {Y.~X.}\ \bibnamefont {Zhao}},\ }\bibfield  {title} {\bibinfo
  {title} {Periodic {C}lifford symmetry algebras on flux lattices},\ }\href
  {https://doi.org/10.1103/physrevb.106.125102} {\bibfield  {journal} {\bibinfo
   {journal} {Phys. Rev. B}\ }\textbf {\bibinfo {volume} {106}},\ \bibinfo
  {pages} {125102} (\bibinfo {year} {2022})}\BibitemShut {NoStop}%
\bibitem [{\citenamefont {Chen}\ \emph {et~al.}(2023)\citenamefont {Chen},
  \citenamefont {Zhang}, \citenamefont {Yang},\ and\ \citenamefont
  {Zhao}}]{Chen2023}%
  \BibitemOpen
  \bibfield  {author} {\bibinfo {author} {\bibfnamefont {Z.~Y.}\ \bibnamefont
  {Chen}}, \bibinfo {author} {\bibfnamefont {Z.}~\bibnamefont {Zhang}},
  \bibinfo {author} {\bibfnamefont {S.~A.}\ \bibnamefont {Yang}},\ and\
  \bibinfo {author} {\bibfnamefont {Y.~X.}\ \bibnamefont {Zhao}},\ }\bibfield
  {title} {\bibinfo {title} {Classification of time-reversal-invariant crystals
  with gauge structures},\ }\href {https://doi.org/10.1038/s41467-023-36447-7}
  {\bibfield  {journal} {\bibinfo  {journal} {Nat. Commun.}\ }\textbf {\bibinfo
  {volume} {14}},\ \bibinfo {pages} {743} (\bibinfo {year} {2023})}\BibitemShut
  {NoStop}%
\bibitem [{\citenamefont {Chiu}\ \emph {et~al.}(2013)\citenamefont {Chiu},
  \citenamefont {Yao},\ and\ \citenamefont {Ryu}}]{Chiu2013}%
  \BibitemOpen
  \bibfield  {author} {\bibinfo {author} {\bibfnamefont {C.-K.}\ \bibnamefont
  {Chiu}}, \bibinfo {author} {\bibfnamefont {H.}~\bibnamefont {Yao}},\ and\
  \bibinfo {author} {\bibfnamefont {S.}~\bibnamefont {Ryu}},\ }\bibfield
  {title} {\bibinfo {title} {Classification of topological insulators and
  superconductors in the presence of reflection symmetry},\ }\href
  {https://doi.org/10.1103/physrevb.88.075142} {\bibfield  {journal} {\bibinfo
  {journal} {Phys. Rev. B}\ }\textbf {\bibinfo {volume} {88}},\ \bibinfo
  {pages} {075142} (\bibinfo {year} {2013})}\BibitemShut {NoStop}%
\bibitem [{Note1()}]{Note1}%
  \BibitemOpen
  \bibinfo {note} {See Supplementary Material for details}\BibitemShut
  {NoStop}%
\bibitem [{\citenamefont {Hou}\ \emph {et~al.}(2007)\citenamefont {Hou},
  \citenamefont {Chamon},\ and\ \citenamefont {Mudry}}]{Hou2007}%
  \BibitemOpen
  \bibfield  {author} {\bibinfo {author} {\bibfnamefont {C.-Y.}\ \bibnamefont
  {Hou}}, \bibinfo {author} {\bibfnamefont {C.}~\bibnamefont {Chamon}},\ and\
  \bibinfo {author} {\bibfnamefont {C.}~\bibnamefont {Mudry}},\ }\bibfield
  {title} {\bibinfo {title} {Electron fractionalization in two-dimensional
  graphenelike structures},\ }\href
  {https://doi.org/10.1103/physrevlett.98.186809} {\bibfield  {journal}
  {\bibinfo  {journal} {Phys. Rev. Lett.}\ }\textbf {\bibinfo {volume} {98}},\
  \bibinfo {pages} {186809} (\bibinfo {year} {2007})}\BibitemShut {NoStop}%
\bibitem [{\citenamefont {Iadecola}\ \emph {et~al.}(2016)\citenamefont
  {Iadecola}, \citenamefont {Schuster},\ and\ \citenamefont
  {Chamon}}]{Iadecola2016}%
  \BibitemOpen
  \bibfield  {author} {\bibinfo {author} {\bibfnamefont {T.}~\bibnamefont
  {Iadecola}}, \bibinfo {author} {\bibfnamefont {T.}~\bibnamefont {Schuster}},\
  and\ \bibinfo {author} {\bibfnamefont {C.}~\bibnamefont {Chamon}},\
  }\bibfield  {title} {\bibinfo {title} {Non-{A}belian braiding of light},\
  }\href {https://doi.org/10.1103/physrevlett.117.073901} {\bibfield  {journal}
  {\bibinfo  {journal} {Phys. Rev. Lett.}\ }\textbf {\bibinfo {volume} {117}},\
  \bibinfo {pages} {073901} (\bibinfo {year} {2016})}\BibitemShut {NoStop}%
\bibitem [{\citenamefont {Gao}\ \emph {et~al.}(2019)\citenamefont {Gao},
  \citenamefont {Torrent}, \citenamefont {Cervera}, \citenamefont {San-Jose},
  \citenamefont {Sánchez-Dehesa},\ and\ \citenamefont
  {Christensen}}]{Gao2019}%
  \BibitemOpen
  \bibfield  {author} {\bibinfo {author} {\bibfnamefont {P.}~\bibnamefont
  {Gao}}, \bibinfo {author} {\bibfnamefont {D.}~\bibnamefont {Torrent}},
  \bibinfo {author} {\bibfnamefont {F.}~\bibnamefont {Cervera}}, \bibinfo
  {author} {\bibfnamefont {P.}~\bibnamefont {San-Jose}}, \bibinfo {author}
  {\bibfnamefont {J.}~\bibnamefont {Sánchez-Dehesa}},\ and\ \bibinfo {author}
  {\bibfnamefont {J.}~\bibnamefont {Christensen}},\ }\bibfield  {title}
  {\bibinfo {title} {Majorana-like zero modes in {K}ekulé distorted sonic
  lattices},\ }\href {https://doi.org/10.1103/physrevlett.123.196601}
  {\bibfield  {journal} {\bibinfo  {journal} {Phys. Rev. Lett.}\ }\textbf
  {\bibinfo {volume} {123}},\ \bibinfo {pages} {196601} (\bibinfo {year}
  {2019})}\BibitemShut {NoStop}%
\bibitem [{\citenamefont {Chen}\ \emph {et~al.}(2019)\citenamefont {Chen},
  \citenamefont {Lera}, \citenamefont {Chaunsali}, \citenamefont {Torrent},
  \citenamefont {Alvarez}, \citenamefont {Yang}, \citenamefont {San‐Jose},\
  and\ \citenamefont {Christensen}}]{Chen2019}%
  \BibitemOpen
  \bibfield  {author} {\bibinfo {author} {\bibfnamefont {C.}~\bibnamefont
  {Chen}}, \bibinfo {author} {\bibfnamefont {N.}~\bibnamefont {Lera}}, \bibinfo
  {author} {\bibfnamefont {R.}~\bibnamefont {Chaunsali}}, \bibinfo {author}
  {\bibfnamefont {D.}~\bibnamefont {Torrent}}, \bibinfo {author} {\bibfnamefont
  {J.~V.}\ \bibnamefont {Alvarez}}, \bibinfo {author} {\bibfnamefont
  {J.}~\bibnamefont {Yang}}, \bibinfo {author} {\bibfnamefont {P.}~\bibnamefont
  {San‐Jose}},\ and\ \bibinfo {author} {\bibfnamefont {J.}~\bibnamefont
  {Christensen}},\ }\bibfield  {title} {\bibinfo {title} {Mechanical analogue
  of a {M}ajorana bound state},\ }\href
  {https://doi.org/10.1002/adma.201904386} {\bibfield  {journal} {\bibinfo
  {journal} {Adv. Mater.}\ }\textbf {\bibinfo {volume} {31}},\ \bibinfo {pages}
  {1904386} (\bibinfo {year} {2019})}\BibitemShut {NoStop}%
\bibitem [{\citenamefont {Noh}\ \emph {et~al.}(2020)\citenamefont {Noh},
  \citenamefont {Schuster}, \citenamefont {Iadecola}, \citenamefont {Huang},
  \citenamefont {Wang}, \citenamefont {Chen}, \citenamefont {Chamon},\ and\
  \citenamefont {Rechtsman}}]{Noh2020}%
  \BibitemOpen
  \bibfield  {author} {\bibinfo {author} {\bibfnamefont {J.}~\bibnamefont
  {Noh}}, \bibinfo {author} {\bibfnamefont {T.}~\bibnamefont {Schuster}},
  \bibinfo {author} {\bibfnamefont {T.}~\bibnamefont {Iadecola}}, \bibinfo
  {author} {\bibfnamefont {S.}~\bibnamefont {Huang}}, \bibinfo {author}
  {\bibfnamefont {M.}~\bibnamefont {Wang}}, \bibinfo {author} {\bibfnamefont
  {K.~P.}\ \bibnamefont {Chen}}, \bibinfo {author} {\bibfnamefont
  {C.}~\bibnamefont {Chamon}},\ and\ \bibinfo {author} {\bibfnamefont {M.~C.}\
  \bibnamefont {Rechtsman}},\ }\bibfield  {title} {\bibinfo {title} {Braiding
  photonic topological zero modes},\ }\href
  {https://doi.org/10.1038/s41567-020-1007-5} {\bibfield  {journal} {\bibinfo
  {journal} {Nat. Phys.}\ }\textbf {\bibinfo {volume} {16}},\ \bibinfo {pages}
  {989} (\bibinfo {year} {2020})}\BibitemShut {NoStop}%
\bibitem [{\citenamefont {Menssen}\ \emph {et~al.}(2020)\citenamefont
  {Menssen}, \citenamefont {Guan}, \citenamefont {Felce}, \citenamefont
  {Booth},\ and\ \citenamefont {Walmsley}}]{Menssen2020}%
  \BibitemOpen
  \bibfield  {author} {\bibinfo {author} {\bibfnamefont {A.~J.}\ \bibnamefont
  {Menssen}}, \bibinfo {author} {\bibfnamefont {J.}~\bibnamefont {Guan}},
  \bibinfo {author} {\bibfnamefont {D.}~\bibnamefont {Felce}}, \bibinfo
  {author} {\bibfnamefont {M.~J.}\ \bibnamefont {Booth}},\ and\ \bibinfo
  {author} {\bibfnamefont {I.~A.}\ \bibnamefont {Walmsley}},\ }\bibfield
  {title} {\bibinfo {title} {Photonic topological mode bound to a vortex},\
  }\href {https://doi.org/10.1103/physrevlett.125.117401} {\bibfield  {journal}
  {\bibinfo  {journal} {Phys. Rev. Lett.}\ }\textbf {\bibinfo {volume} {125}},\
  \bibinfo {pages} {117401} (\bibinfo {year} {2020})}\BibitemShut {NoStop}%
\bibitem [{\citenamefont {Gao}\ \emph {et~al.}(2020)\citenamefont {Gao},
  \citenamefont {Yang}, \citenamefont {Lin}, \citenamefont {Zhang},
  \citenamefont {Li}, \citenamefont {Bo}, \citenamefont {Wang},\ and\
  \citenamefont {Lu}}]{Gao2020}%
  \BibitemOpen
  \bibfield  {author} {\bibinfo {author} {\bibfnamefont {X.}~\bibnamefont
  {Gao}}, \bibinfo {author} {\bibfnamefont {L.}~\bibnamefont {Yang}}, \bibinfo
  {author} {\bibfnamefont {H.}~\bibnamefont {Lin}}, \bibinfo {author}
  {\bibfnamefont {L.}~\bibnamefont {Zhang}}, \bibinfo {author} {\bibfnamefont
  {J.}~\bibnamefont {Li}}, \bibinfo {author} {\bibfnamefont {F.}~\bibnamefont
  {Bo}}, \bibinfo {author} {\bibfnamefont {Z.}~\bibnamefont {Wang}},\ and\
  \bibinfo {author} {\bibfnamefont {L.}~\bibnamefont {Lu}},\ }\bibfield
  {title} {\bibinfo {title} {Dirac-vortex topological cavities},\ }\href
  {https://doi.org/10.1038/s41565-020-0773-7} {\bibfield  {journal} {\bibinfo
  {journal} {Nat. Nanotechnol.}\ }\textbf {\bibinfo {volume} {15}},\ \bibinfo
  {pages} {1012} (\bibinfo {year} {2020})}\BibitemShut {NoStop}%
\bibitem [{\citenamefont {Lin}\ and\ \citenamefont {Lu}(2020)}]{Lin2020}%
  \BibitemOpen
  \bibfield  {author} {\bibinfo {author} {\bibfnamefont {H.}~\bibnamefont
  {Lin}}\ and\ \bibinfo {author} {\bibfnamefont {L.}~\bibnamefont {Lu}},\
  }\bibfield  {title} {\bibinfo {title} {Dirac-vortex topological photonic
  crystal fibre},\ }\href {https://doi.org/10.1038/s41377-020-00432-2}
  {\bibfield  {journal} {\bibinfo  {journal} {Light Sci. Appl.}\ }\textbf
  {\bibinfo {volume} {9}},\ \bibinfo {pages} {202} (\bibinfo {year}
  {2020})}\BibitemShut {NoStop}%
\bibitem [{\citenamefont {Cheng}\ \emph {et~al.}(2024)\citenamefont {Cheng},
  \citenamefont {Yang}, \citenamefont {Wang},\ and\ \citenamefont
  {Lu}}]{Cheng2024}%
  \BibitemOpen
  \bibfield  {author} {\bibinfo {author} {\bibfnamefont {H.}~\bibnamefont
  {Cheng}}, \bibinfo {author} {\bibfnamefont {J.}~\bibnamefont {Yang}},
  \bibinfo {author} {\bibfnamefont {Z.}~\bibnamefont {Wang}},\ and\ \bibinfo
  {author} {\bibfnamefont {L.}~\bibnamefont {Lu}},\ }\bibfield  {title}
  {\bibinfo {title} {Observation of monopole topological mode},\ }\href
  {https://doi.org/10.1038/s41467-024-51670-6} {\bibfield  {journal} {\bibinfo
  {journal} {Nat. Commun.}\ }\textbf {\bibinfo {volume} {15}},\ \bibinfo
  {pages} {7327} (\bibinfo {year} {2024})}\BibitemShut {NoStop}%
\bibitem [{\citenamefont {Teo}\ and\ \citenamefont
  {Kane}(2010)}]{teo2010topological}%
  \BibitemOpen
  \bibfield  {author} {\bibinfo {author} {\bibfnamefont {J.~C.~Y.}\
  \bibnamefont {Teo}}\ and\ \bibinfo {author} {\bibfnamefont {C.~L.}\
  \bibnamefont {Kane}},\ }\bibfield  {title} {\bibinfo {title} {Topological
  defects and gapless modes in insulators and superconductors},\ }\href
  {https://doi.org/10.1103/PhysRevB.82.115120} {\bibfield  {journal} {\bibinfo
  {journal} {Phys. Rev. B}\ }\textbf {\bibinfo {volume} {82}},\ \bibinfo
  {pages} {115120} (\bibinfo {year} {2010})}\BibitemShut {NoStop}%
\bibitem [{\citenamefont {Keil}\ \emph {et~al.}(2016)\citenamefont {Keil},
  \citenamefont {Poli}, \citenamefont {Heinrich}, \citenamefont {Arkinstall},
  \citenamefont {Weihs}, \citenamefont {Schomerus},\ and\ \citenamefont
  {Szameit}}]{Keil2016}%
  \BibitemOpen
  \bibfield  {author} {\bibinfo {author} {\bibfnamefont {R.}~\bibnamefont
  {Keil}}, \bibinfo {author} {\bibfnamefont {C.}~\bibnamefont {Poli}}, \bibinfo
  {author} {\bibfnamefont {M.}~\bibnamefont {Heinrich}}, \bibinfo {author}
  {\bibfnamefont {J.}~\bibnamefont {Arkinstall}}, \bibinfo {author}
  {\bibfnamefont {G.}~\bibnamefont {Weihs}}, \bibinfo {author} {\bibfnamefont
  {H.}~\bibnamefont {Schomerus}},\ and\ \bibinfo {author} {\bibfnamefont
  {A.}~\bibnamefont {Szameit}},\ }\bibfield  {title} {\bibinfo {title}
  {Universal sign control of coupling in tight-binding lattices},\ }\href
  {https://doi.org/10.1103/physrevlett.116.213901} {\bibfield  {journal}
  {\bibinfo  {journal} {Phys. Rev. Lett.}\ }\textbf {\bibinfo {volume} {116}},\
  \bibinfo {pages} {213901} (\bibinfo {year} {2016})}\BibitemShut {NoStop}%
\bibitem [{\citenamefont {Imhof}\ \emph {et~al.}(2018)\citenamefont {Imhof},
  \citenamefont {Berger}, \citenamefont {Bayer}, \citenamefont {Brehm},
  \citenamefont {Molenkamp}, \citenamefont {Kiessling}, \citenamefont
  {Schindler}, \citenamefont {Lee}, \citenamefont {Greiter}, \citenamefont
  {Neupert},\ and\ \citenamefont {Thomale}}]{Imhof2018}%
  \BibitemOpen
  \bibfield  {author} {\bibinfo {author} {\bibfnamefont {S.}~\bibnamefont
  {Imhof}}, \bibinfo {author} {\bibfnamefont {C.}~\bibnamefont {Berger}},
  \bibinfo {author} {\bibfnamefont {F.}~\bibnamefont {Bayer}}, \bibinfo
  {author} {\bibfnamefont {J.}~\bibnamefont {Brehm}}, \bibinfo {author}
  {\bibfnamefont {L.~W.}\ \bibnamefont {Molenkamp}}, \bibinfo {author}
  {\bibfnamefont {T.}~\bibnamefont {Kiessling}}, \bibinfo {author}
  {\bibfnamefont {F.}~\bibnamefont {Schindler}}, \bibinfo {author}
  {\bibfnamefont {C.~H.}\ \bibnamefont {Lee}}, \bibinfo {author} {\bibfnamefont
  {M.}~\bibnamefont {Greiter}}, \bibinfo {author} {\bibfnamefont
  {T.}~\bibnamefont {Neupert}},\ and\ \bibinfo {author} {\bibfnamefont
  {R.}~\bibnamefont {Thomale}},\ }\bibfield  {title} {\bibinfo {title}
  {Topolectrical-circuit realization of topological corner modes},\ }\href
  {https://doi.org/10.1038/s41567-018-0246-1} {\bibfield  {journal} {\bibinfo
  {journal} {Nat. Phys.}\ }\textbf {\bibinfo {volume} {14}},\ \bibinfo {pages}
  {925} (\bibinfo {year} {2018})}\BibitemShut {NoStop}%
\bibitem [{\citenamefont {Lee}\ \emph {et~al.}(2018)\citenamefont {Lee},
  \citenamefont {Imhof}, \citenamefont {Berger}, \citenamefont {Bayer},
  \citenamefont {Brehm}, \citenamefont {Molenkamp}, \citenamefont {Kiessling},\
  and\ \citenamefont {Thomale}}]{Lee2018}%
  \BibitemOpen
  \bibfield  {author} {\bibinfo {author} {\bibfnamefont {C.~H.}\ \bibnamefont
  {Lee}}, \bibinfo {author} {\bibfnamefont {S.}~\bibnamefont {Imhof}}, \bibinfo
  {author} {\bibfnamefont {C.}~\bibnamefont {Berger}}, \bibinfo {author}
  {\bibfnamefont {F.}~\bibnamefont {Bayer}}, \bibinfo {author} {\bibfnamefont
  {J.}~\bibnamefont {Brehm}}, \bibinfo {author} {\bibfnamefont {L.~W.}\
  \bibnamefont {Molenkamp}}, \bibinfo {author} {\bibfnamefont {T.}~\bibnamefont
  {Kiessling}},\ and\ \bibinfo {author} {\bibfnamefont {R.}~\bibnamefont
  {Thomale}},\ }\bibfield  {title} {\bibinfo {title} {Topolectrical circuits},\
  }\href {https://doi.org/10.1038/s42005-018-0035-2} {\bibfield  {journal}
  {\bibinfo  {journal} {Commun. Phys.}\ }\textbf {\bibinfo {volume} {1}},\
  \bibinfo {pages} {39} (\bibinfo {year} {2018})}\BibitemShut {NoStop}%
\bibitem [{\citenamefont {Bao}\ \emph {et~al.}(2019)\citenamefont {Bao},
  \citenamefont {Zou}, \citenamefont {Zhang}, \citenamefont {He}, \citenamefont
  {Sun},\ and\ \citenamefont {Zhang}}]{Bao2019}%
  \BibitemOpen
  \bibfield  {author} {\bibinfo {author} {\bibfnamefont {J.}~\bibnamefont
  {Bao}}, \bibinfo {author} {\bibfnamefont {D.}~\bibnamefont {Zou}}, \bibinfo
  {author} {\bibfnamefont {W.}~\bibnamefont {Zhang}}, \bibinfo {author}
  {\bibfnamefont {W.}~\bibnamefont {He}}, \bibinfo {author} {\bibfnamefont
  {H.}~\bibnamefont {Sun}},\ and\ \bibinfo {author} {\bibfnamefont
  {X.}~\bibnamefont {Zhang}},\ }\bibfield  {title} {\bibinfo {title}
  {Topoelectrical circuit octupole insulator with topologically protected
  corner states},\ }\href {https://doi.org/10.1103/physrevb.100.201406}
  {\bibfield  {journal} {\bibinfo  {journal} {Phys. Rev. B}\ }\textbf {\bibinfo
  {volume} {100}},\ \bibinfo {pages} {201406(R)} (\bibinfo {year}
  {2019})}\BibitemShut {NoStop}%
\bibitem [{\citenamefont {Liu}\ \emph {et~al.}(2020)\citenamefont {Liu},
  \citenamefont {Ma}, \citenamefont {Zhang}, \citenamefont {Zhang},
  \citenamefont {Yang}, \citenamefont {You}, \citenamefont {Gao}, \citenamefont
  {Xiang}, \citenamefont {Cui},\ and\ \citenamefont {Zhang}}]{Liu2020}%
  \BibitemOpen
  \bibfield  {author} {\bibinfo {author} {\bibfnamefont {S.}~\bibnamefont
  {Liu}}, \bibinfo {author} {\bibfnamefont {S.}~\bibnamefont {Ma}}, \bibinfo
  {author} {\bibfnamefont {Q.}~\bibnamefont {Zhang}}, \bibinfo {author}
  {\bibfnamefont {L.}~\bibnamefont {Zhang}}, \bibinfo {author} {\bibfnamefont
  {C.}~\bibnamefont {Yang}}, \bibinfo {author} {\bibfnamefont {O.}~\bibnamefont
  {You}}, \bibinfo {author} {\bibfnamefont {W.}~\bibnamefont {Gao}}, \bibinfo
  {author} {\bibfnamefont {Y.}~\bibnamefont {Xiang}}, \bibinfo {author}
  {\bibfnamefont {T.~J.}\ \bibnamefont {Cui}},\ and\ \bibinfo {author}
  {\bibfnamefont {S.}~\bibnamefont {Zhang}},\ }\bibfield  {title} {\bibinfo
  {title} {Octupole corner state in a three-dimensional topological circuit},\
  }\href {https://doi.org/10.1038/s41377-020-00381-w} {\bibfield  {journal}
  {\bibinfo  {journal} {Light Sci. Appl.}\ }\textbf {\bibinfo {volume} {9}},\
  \bibinfo {pages} {145} (\bibinfo {year} {2020})}\BibitemShut {NoStop}%
\bibitem [{\citenamefont {Yamada}\ \emph {et~al.}(2022)\citenamefont {Yamada},
  \citenamefont {Li}, \citenamefont {Lin}, \citenamefont {Peterson},
  \citenamefont {Hughes},\ and\ \citenamefont {Bahl}}]{Yamada2022}%
  \BibitemOpen
  \bibfield  {author} {\bibinfo {author} {\bibfnamefont {S.~S.}\ \bibnamefont
  {Yamada}}, \bibinfo {author} {\bibfnamefont {T.}~\bibnamefont {Li}}, \bibinfo
  {author} {\bibfnamefont {M.}~\bibnamefont {Lin}}, \bibinfo {author}
  {\bibfnamefont {C.~W.}\ \bibnamefont {Peterson}}, \bibinfo {author}
  {\bibfnamefont {T.~L.}\ \bibnamefont {Hughes}},\ and\ \bibinfo {author}
  {\bibfnamefont {G.}~\bibnamefont {Bahl}},\ }\bibfield  {title} {\bibinfo
  {title} {Bound states at partial dislocation defects in multipole
  higher-order topological insulators},\ }\href
  {https://doi.org/10.1038/s41467-022-29785-5} {\bibfield  {journal} {\bibinfo
  {journal} {Nat. Commun.}\ }\textbf {\bibinfo {volume} {13}},\ \bibinfo
  {pages} {2035} (\bibinfo {year} {2022})}\BibitemShut {NoStop}%
\bibitem [{\citenamefont {Liu}\ \emph {et~al.}(2019)\citenamefont {Liu},
  \citenamefont {Zhang}, \citenamefont {Ai}, \citenamefont {Gong},
  \citenamefont {Kawabata}, \citenamefont {Ueda},\ and\ \citenamefont
  {Nori}}]{Liu2019}%
  \BibitemOpen
  \bibfield  {author} {\bibinfo {author} {\bibfnamefont {T.}~\bibnamefont
  {Liu}}, \bibinfo {author} {\bibfnamefont {Y.-R.}\ \bibnamefont {Zhang}},
  \bibinfo {author} {\bibfnamefont {Q.}~\bibnamefont {Ai}}, \bibinfo {author}
  {\bibfnamefont {Z.}~\bibnamefont {Gong}}, \bibinfo {author} {\bibfnamefont
  {K.}~\bibnamefont {Kawabata}}, \bibinfo {author} {\bibfnamefont
  {M.}~\bibnamefont {Ueda}},\ and\ \bibinfo {author} {\bibfnamefont
  {F.}~\bibnamefont {Nori}},\ }\bibfield  {title} {\bibinfo {title}
  {Second-order topological phases in non-{H}ermitian systems},\ }\href
  {https://doi.org/10.1103/physrevlett.122.076801} {\bibfield  {journal}
  {\bibinfo  {journal} {Phys. Rev. Lett.}\ }\textbf {\bibinfo {volume} {122}},\
  \bibinfo {pages} {076801} (\bibinfo {year} {2019})}\BibitemShut {NoStop}%
\bibitem [{\citenamefont {Lee}\ \emph {et~al.}(2019)\citenamefont {Lee},
  \citenamefont {Li},\ and\ \citenamefont {Gong}}]{Lee2019}%
  \BibitemOpen
  \bibfield  {author} {\bibinfo {author} {\bibfnamefont {C.~H.}\ \bibnamefont
  {Lee}}, \bibinfo {author} {\bibfnamefont {L.}~\bibnamefont {Li}},\ and\
  \bibinfo {author} {\bibfnamefont {J.}~\bibnamefont {Gong}},\ }\bibfield
  {title} {\bibinfo {title} {Hybrid higher-order skin-topological modes in
  nonreciprocal systems},\ }\href
  {https://doi.org/10.1103/physrevlett.123.016805} {\bibfield  {journal}
  {\bibinfo  {journal} {Phys. Rev. Lett.}\ }\textbf {\bibinfo {volume} {123}},\
  \bibinfo {pages} {016805} (\bibinfo {year} {2019})}\BibitemShut {NoStop}%
\bibitem [{\citenamefont {Luo}\ and\ \citenamefont {Zhang}(2019)}]{Luo2019}%
  \BibitemOpen
  \bibfield  {author} {\bibinfo {author} {\bibfnamefont {X.-W.}\ \bibnamefont
  {Luo}}\ and\ \bibinfo {author} {\bibfnamefont {C.}~\bibnamefont {Zhang}},\
  }\bibfield  {title} {\bibinfo {title} {Higher-order topological corner states
  induced by gain and loss},\ }\href
  {https://doi.org/10.1103/physrevlett.123.073601} {\bibfield  {journal}
  {\bibinfo  {journal} {Phys. Rev. Lett.}\ }\textbf {\bibinfo {volume} {123}},\
  \bibinfo {pages} {073601} (\bibinfo {year} {2019})}\BibitemShut {NoStop}%
\bibitem [{\citenamefont {Kawabata}\ \emph
  {et~al.}(2019{\natexlab{a}})\citenamefont {Kawabata}, \citenamefont
  {Bessho},\ and\ \citenamefont {Sato}}]{Kawabata2019}%
  \BibitemOpen
  \bibfield  {author} {\bibinfo {author} {\bibfnamefont {K.}~\bibnamefont
  {Kawabata}}, \bibinfo {author} {\bibfnamefont {T.}~\bibnamefont {Bessho}},\
  and\ \bibinfo {author} {\bibfnamefont {M.}~\bibnamefont {Sato}},\ }\bibfield
  {title} {\bibinfo {title} {Classification of exceptional points and
  non-{H}ermitian topological semimetals},\ }\href
  {https://doi.org/10.1103/physrevlett.123.066405} {\bibfield  {journal}
  {\bibinfo  {journal} {Phys. Rev. Lett.}\ }\textbf {\bibinfo {volume} {123}},\
  \bibinfo {pages} {066405} (\bibinfo {year} {2019}{\natexlab{a}})}\BibitemShut
  {NoStop}%
\bibitem [{\citenamefont {Kawabata}\ \emph
  {et~al.}(2019{\natexlab{b}})\citenamefont {Kawabata}, \citenamefont
  {Shiozaki}, \citenamefont {Ueda},\ and\ \citenamefont
  {Sato}}]{Kawabata2019a}%
  \BibitemOpen
  \bibfield  {author} {\bibinfo {author} {\bibfnamefont {K.}~\bibnamefont
  {Kawabata}}, \bibinfo {author} {\bibfnamefont {K.}~\bibnamefont {Shiozaki}},
  \bibinfo {author} {\bibfnamefont {M.}~\bibnamefont {Ueda}},\ and\ \bibinfo
  {author} {\bibfnamefont {M.}~\bibnamefont {Sato}},\ }\bibfield  {title}
  {\bibinfo {title} {Symmetry and topology in non-{H}ermitian physics},\ }\href
  {https://doi.org/10.1103/physrevx.9.041015} {\bibfield  {journal} {\bibinfo
  {journal} {Phys. Rev. X}\ }\textbf {\bibinfo {volume} {9}},\ \bibinfo {pages}
  {041015} (\bibinfo {year} {2019}{\natexlab{b}})}\BibitemShut {NoStop}%
\bibitem [{\citenamefont {Gao}\ \emph {et~al.}(2021)\citenamefont {Gao},
  \citenamefont {Xue}, \citenamefont {Gu}, \citenamefont {Liu}, \citenamefont
  {Zhu},\ and\ \citenamefont {Zhang}}]{Gao2021}%
  \BibitemOpen
  \bibfield  {author} {\bibinfo {author} {\bibfnamefont {H.}~\bibnamefont
  {Gao}}, \bibinfo {author} {\bibfnamefont {H.}~\bibnamefont {Xue}}, \bibinfo
  {author} {\bibfnamefont {Z.}~\bibnamefont {Gu}}, \bibinfo {author}
  {\bibfnamefont {T.}~\bibnamefont {Liu}}, \bibinfo {author} {\bibfnamefont
  {J.}~\bibnamefont {Zhu}},\ and\ \bibinfo {author} {\bibfnamefont
  {B.}~\bibnamefont {Zhang}},\ }\bibfield  {title} {\bibinfo {title}
  {Non-{H}ermitian route to higher-order topology in an acoustic crystal},\
  }\href {https://doi.org/10.1038/s41467-021-22223-y} {\bibfield  {journal}
  {\bibinfo  {journal} {Nat. Commun.}\ }\textbf {\bibinfo {volume} {12}},\
  \bibinfo {pages} {1888} (\bibinfo {year} {2021})}\BibitemShut {NoStop}%
\bibitem [{\citenamefont {Serra-Garcia}\ \emph {et~al.}(2019)\citenamefont
  {Serra-Garcia}, \citenamefont {Süsstrunk},\ and\ \citenamefont
  {Huber}}]{SerraGarcia2019}%
  \BibitemOpen
  \bibfield  {author} {\bibinfo {author} {\bibfnamefont {M.}~\bibnamefont
  {Serra-Garcia}}, \bibinfo {author} {\bibfnamefont {R.}~\bibnamefont
  {Süsstrunk}},\ and\ \bibinfo {author} {\bibfnamefont {S.~D.}\ \bibnamefont
  {Huber}},\ }\bibfield  {title} {\bibinfo {title} {Observation of quadrupole
  transitions and edge mode topology in an {LC} circuit network},\ }\href
  {https://doi.org/10.1103/physrevb.99.020304} {\bibfield  {journal} {\bibinfo
  {journal} {Phys. Rev. B}\ }\textbf {\bibinfo {volume} {99}},\ \bibinfo
  {pages} {020304(R)} (\bibinfo {year} {2019})}\BibitemShut {NoStop}%
\bibitem [{\citenamefont {Zangeneh-Nejad}\ and\ \citenamefont
  {Fleury}(2019)}]{ZangenehNejad2019}%
  \BibitemOpen
  \bibfield  {author} {\bibinfo {author} {\bibfnamefont {F.}~\bibnamefont
  {Zangeneh-Nejad}}\ and\ \bibinfo {author} {\bibfnamefont {R.}~\bibnamefont
  {Fleury}},\ }\bibfield  {title} {\bibinfo {title} {Nonlinear second-order
  topological insulators},\ }\href
  {https://doi.org/10.1103/physrevlett.123.053902} {\bibfield  {journal}
  {\bibinfo  {journal} {Phys. Rev. Lett.}\ }\textbf {\bibinfo {volume} {123}},\
  \bibinfo {pages} {053902} (\bibinfo {year} {2019})}\BibitemShut {NoStop}%
\bibitem [{\citenamefont {Wang}\ \emph {et~al.}(2019)\citenamefont {Wang},
  \citenamefont {Lang}, \citenamefont {Lee}, \citenamefont {Zhang},\ and\
  \citenamefont {Chong}}]{Wang2019a}%
  \BibitemOpen
  \bibfield  {author} {\bibinfo {author} {\bibfnamefont {Y.}~\bibnamefont
  {Wang}}, \bibinfo {author} {\bibfnamefont {L.-J.}\ \bibnamefont {Lang}},
  \bibinfo {author} {\bibfnamefont {C.~H.}\ \bibnamefont {Lee}}, \bibinfo
  {author} {\bibfnamefont {B.}~\bibnamefont {Zhang}},\ and\ \bibinfo {author}
  {\bibfnamefont {Y.~D.}\ \bibnamefont {Chong}},\ }\bibfield  {title} {\bibinfo
  {title} {Topologically enhanced harmonic generation in a nonlinear
  transmission line metamaterial},\ }\href
  {https://doi.org/10.1038/s41467-019-08966-9} {\bibfield  {journal} {\bibinfo
  {journal} {Nat. Commun.}\ }\textbf {\bibinfo {volume} {10}},\ \bibinfo
  {pages} {1102} (\bibinfo {year} {2019})}\BibitemShut {NoStop}%
\bibitem [{\citenamefont {Kirsch}\ \emph {et~al.}(2021)\citenamefont {Kirsch},
  \citenamefont {Zhang}, \citenamefont {Kremer}, \citenamefont {Maczewsky},
  \citenamefont {Ivanov}, \citenamefont {Kartashov}, \citenamefont {Torner},
  \citenamefont {Bauer}, \citenamefont {Szameit},\ and\ \citenamefont
  {Heinrich}}]{Kirsch2021}%
  \BibitemOpen
  \bibfield  {author} {\bibinfo {author} {\bibfnamefont {M.~S.}\ \bibnamefont
  {Kirsch}}, \bibinfo {author} {\bibfnamefont {Y.}~\bibnamefont {Zhang}},
  \bibinfo {author} {\bibfnamefont {M.}~\bibnamefont {Kremer}}, \bibinfo
  {author} {\bibfnamefont {L.~J.}\ \bibnamefont {Maczewsky}}, \bibinfo {author}
  {\bibfnamefont {S.~K.}\ \bibnamefont {Ivanov}}, \bibinfo {author}
  {\bibfnamefont {Y.~V.}\ \bibnamefont {Kartashov}}, \bibinfo {author}
  {\bibfnamefont {L.}~\bibnamefont {Torner}}, \bibinfo {author} {\bibfnamefont
  {D.}~\bibnamefont {Bauer}}, \bibinfo {author} {\bibfnamefont
  {A.}~\bibnamefont {Szameit}},\ and\ \bibinfo {author} {\bibfnamefont
  {M.}~\bibnamefont {Heinrich}},\ }\bibfield  {title} {\bibinfo {title}
  {Nonlinear second-order photonic topological insulators},\ }\href
  {https://doi.org/10.1038/s41567-021-01275-3} {\bibfield  {journal} {\bibinfo
  {journal} {Nat. Phys.}\ }\textbf {\bibinfo {volume} {17}},\ \bibinfo {pages}
  {995} (\bibinfo {year} {2021})}\BibitemShut {NoStop}%
\bibitem [{\citenamefont {Lu}\ \emph {et~al.}(2018)\citenamefont {Lu},
  \citenamefont {Gao},\ and\ \citenamefont {Wang}}]{Lu2018}%
  \BibitemOpen
  \bibfield  {author} {\bibinfo {author} {\bibfnamefont {L.}~\bibnamefont
  {Lu}}, \bibinfo {author} {\bibfnamefont {H.}~\bibnamefont {Gao}},\ and\
  \bibinfo {author} {\bibfnamefont {Z.}~\bibnamefont {Wang}},\ }\bibfield
  {title} {\bibinfo {title} {Topological one-way fiber of second {C}hern
  number},\ }\href {https://doi.org/10.1038/s41467-018-07817-3} {\bibfield
  {journal} {\bibinfo  {journal} {Nat. Commun.}\ }\textbf {\bibinfo {volume}
  {9}},\ \bibinfo {pages} {5384} (\bibinfo {year} {2018})}\BibitemShut
  {NoStop}%
\bibitem [{\citenamefont {Queiroz}\ \emph {et~al.}(2019)\citenamefont
  {Queiroz}, \citenamefont {Fulga}, \citenamefont {Avraham}, \citenamefont
  {Beidenkopf},\ and\ \citenamefont {Cano}}]{Queiroz2019}%
  \BibitemOpen
  \bibfield  {author} {\bibinfo {author} {\bibfnamefont {R.}~\bibnamefont
  {Queiroz}}, \bibinfo {author} {\bibfnamefont {I.~C.}\ \bibnamefont {Fulga}},
  \bibinfo {author} {\bibfnamefont {N.}~\bibnamefont {Avraham}}, \bibinfo
  {author} {\bibfnamefont {H.}~\bibnamefont {Beidenkopf}},\ and\ \bibinfo
  {author} {\bibfnamefont {J.}~\bibnamefont {Cano}},\ }\bibfield  {title}
  {\bibinfo {title} {Partial lattice defects in higher-order topological
  insulators},\ }\href {https://doi.org/10.1103/physrevlett.123.266802}
  {\bibfield  {journal} {\bibinfo  {journal} {Phys. Rev. Lett.}\ }\textbf
  {\bibinfo {volume} {123}},\ \bibinfo {pages} {266802} (\bibinfo {year}
  {2019})}\BibitemShut {NoStop}%
\bibitem [{\citenamefont {Ota}\ \emph {et~al.}(2018)\citenamefont {Ota},
  \citenamefont {Katsumi}, \citenamefont {Watanabe}, \citenamefont {Iwamoto},\
  and\ \citenamefont {Arakawa}}]{Ota2018}%
  \BibitemOpen
  \bibfield  {author} {\bibinfo {author} {\bibfnamefont {Y.}~\bibnamefont
  {Ota}}, \bibinfo {author} {\bibfnamefont {R.}~\bibnamefont {Katsumi}},
  \bibinfo {author} {\bibfnamefont {K.}~\bibnamefont {Watanabe}}, \bibinfo
  {author} {\bibfnamefont {S.}~\bibnamefont {Iwamoto}},\ and\ \bibinfo {author}
  {\bibfnamefont {Y.}~\bibnamefont {Arakawa}},\ }\bibfield  {title} {\bibinfo
  {title} {Topological photonic crystal nanocavity laser},\ }\href
  {https://doi.org/10.1038/s42005-018-0083-7} {\bibfield  {journal} {\bibinfo
  {journal} {Commun. Phys.}\ }\textbf {\bibinfo {volume} {1}},\ \bibinfo
  {pages} {86} (\bibinfo {year} {2018})}\BibitemShut {NoStop}%
\bibitem [{\citenamefont {Kim}\ \emph {et~al.}(2020)\citenamefont {Kim},
  \citenamefont {Hwang}, \citenamefont {Smirnova}, \citenamefont {Jeong},
  \citenamefont {Kivshar},\ and\ \citenamefont {Park}}]{Kim2020}%
  \BibitemOpen
  \bibfield  {author} {\bibinfo {author} {\bibfnamefont {H.-R.}\ \bibnamefont
  {Kim}}, \bibinfo {author} {\bibfnamefont {M.-S.}\ \bibnamefont {Hwang}},
  \bibinfo {author} {\bibfnamefont {D.}~\bibnamefont {Smirnova}}, \bibinfo
  {author} {\bibfnamefont {K.-Y.}\ \bibnamefont {Jeong}}, \bibinfo {author}
  {\bibfnamefont {Y.}~\bibnamefont {Kivshar}},\ and\ \bibinfo {author}
  {\bibfnamefont {H.-G.}\ \bibnamefont {Park}},\ }\bibfield  {title} {\bibinfo
  {title} {Multipolar lasing modes from topological corner states},\ }\href
  {https://doi.org/10.1038/s41467-020-19609-9} {\bibfield  {journal} {\bibinfo
  {journal} {Nat. Commun.}\ }\textbf {\bibinfo {volume} {11}},\ \bibinfo
  {pages} {5758} (\bibinfo {year} {2020})}\BibitemShut {NoStop}%
\bibitem [{\citenamefont {Smirnova}\ \emph {et~al.}(2020)\citenamefont
  {Smirnova}, \citenamefont {Tripathi}, \citenamefont {Kruk}, \citenamefont
  {Hwang}, \citenamefont {Kim}, \citenamefont {Park},\ and\ \citenamefont
  {Kivshar}}]{Smirnova2020}%
  \BibitemOpen
  \bibfield  {author} {\bibinfo {author} {\bibfnamefont {D.}~\bibnamefont
  {Smirnova}}, \bibinfo {author} {\bibfnamefont {A.}~\bibnamefont {Tripathi}},
  \bibinfo {author} {\bibfnamefont {S.}~\bibnamefont {Kruk}}, \bibinfo {author}
  {\bibfnamefont {M.-S.}\ \bibnamefont {Hwang}}, \bibinfo {author}
  {\bibfnamefont {H.-R.}\ \bibnamefont {Kim}}, \bibinfo {author} {\bibfnamefont
  {H.-G.}\ \bibnamefont {Park}},\ and\ \bibinfo {author} {\bibfnamefont
  {Y.}~\bibnamefont {Kivshar}},\ }\bibfield  {title} {\bibinfo {title}
  {Room-temperature lasing from nanophotonic topological cavities},\ }\href
  {https://doi.org/10.1038/s41377-020-00350-3} {\bibfield  {journal} {\bibinfo
  {journal} {Light Sci. Appl.}\ }\textbf {\bibinfo {volume} {9}},\ \bibinfo
  {pages} {127} (\bibinfo {year} {2020})}\BibitemShut {NoStop}%
\bibitem [{\citenamefont {Laforge}\ \emph {et~al.}(2021)\citenamefont
  {Laforge}, \citenamefont {Wiltshaw}, \citenamefont {Craster}, \citenamefont
  {Laude}, \citenamefont {Mart{\'{\i}}nez}, \citenamefont {Dupont},
  \citenamefont {Guenneau}, \citenamefont {Kadic},\ and\ \citenamefont
  {Makwana}}]{Laforge2021}%
  \BibitemOpen
  \bibfield  {author} {\bibinfo {author} {\bibfnamefont {N.}~\bibnamefont
  {Laforge}}, \bibinfo {author} {\bibfnamefont {R.}~\bibnamefont {Wiltshaw}},
  \bibinfo {author} {\bibfnamefont {R.~V.}\ \bibnamefont {Craster}}, \bibinfo
  {author} {\bibfnamefont {V.}~\bibnamefont {Laude}}, \bibinfo {author}
  {\bibfnamefont {J.~A.~I.}\ \bibnamefont {Mart{\'{\i}}nez}}, \bibinfo {author}
  {\bibfnamefont {G.}~\bibnamefont {Dupont}}, \bibinfo {author} {\bibfnamefont
  {S.}~\bibnamefont {Guenneau}}, \bibinfo {author} {\bibfnamefont
  {M.}~\bibnamefont {Kadic}},\ and\ \bibinfo {author} {\bibfnamefont {M.~P.}\
  \bibnamefont {Makwana}},\ }\bibfield  {title} {\bibinfo {title} {Acoustic
  topological circuitry in square and rectangular phononic crystals},\ }\href
  {https://doi.org/10.1103/physrevapplied.15.054056} {\bibfield  {journal}
  {\bibinfo  {journal} {Phys. Rev. Appl.}\ }\textbf {\bibinfo {volume} {15}},\
  \bibinfo {pages} {054056} (\bibinfo {year} {2021})}\BibitemShut {NoStop}%
\bibitem [{\citenamefont {Lin}\ \emph {et~al.}(2023)\citenamefont {Lin},
  \citenamefont {Wang}, \citenamefont {Liu}, \citenamefont {Xue}, \citenamefont
  {Zhang}, \citenamefont {Chong},\ and\ \citenamefont {Jiang}}]{Lin2023}%
  \BibitemOpen
  \bibfield  {author} {\bibinfo {author} {\bibfnamefont {Z.-K.}\ \bibnamefont
  {Lin}}, \bibinfo {author} {\bibfnamefont {Q.}~\bibnamefont {Wang}}, \bibinfo
  {author} {\bibfnamefont {Y.}~\bibnamefont {Liu}}, \bibinfo {author}
  {\bibfnamefont {H.}~\bibnamefont {Xue}}, \bibinfo {author} {\bibfnamefont
  {B.}~\bibnamefont {Zhang}}, \bibinfo {author} {\bibfnamefont
  {Y.}~\bibnamefont {Chong}},\ and\ \bibinfo {author} {\bibfnamefont {J.-H.}\
  \bibnamefont {Jiang}},\ }\bibfield  {title} {\bibinfo {title} {Topological
  phenomena at defects in acoustic, photonic and solid-state lattices},\ }\href
  {https://doi.org/10.1038/s42254-023-00602-2} {\bibfield  {journal} {\bibinfo
  {journal} {Nat. Rev. Phys.}\ }\textbf {\bibinfo {volume} {5}},\ \bibinfo
  {pages} {483} (\bibinfo {year} {2023})}\BibitemShut {NoStop}%
\bibitem [{\citenamefont {Ma}\ \emph {et~al.}(2024)\citenamefont {Ma},
  \citenamefont {Jia}, \citenamefont {Zhang}, \citenamefont {Guan},
  \citenamefont {Ge}, \citenamefont {Sun}, \citenamefont {Yuan}, \citenamefont
  {Chen}, \citenamefont {Yang},\ and\ \citenamefont {Zhang}}]{Ma2024}%
  \BibitemOpen
  \bibfield  {author} {\bibinfo {author} {\bibfnamefont {J.}~\bibnamefont
  {Ma}}, \bibinfo {author} {\bibfnamefont {D.}~\bibnamefont {Jia}}, \bibinfo
  {author} {\bibfnamefont {L.}~\bibnamefont {Zhang}}, \bibinfo {author}
  {\bibfnamefont {Y.-j.}\ \bibnamefont {Guan}}, \bibinfo {author}
  {\bibfnamefont {Y.}~\bibnamefont {Ge}}, \bibinfo {author} {\bibfnamefont
  {H.-x.}\ \bibnamefont {Sun}}, \bibinfo {author} {\bibfnamefont {S.-q.}\
  \bibnamefont {Yuan}}, \bibinfo {author} {\bibfnamefont {H.}~\bibnamefont
  {Chen}}, \bibinfo {author} {\bibfnamefont {Y.}~\bibnamefont {Yang}},\ and\
  \bibinfo {author} {\bibfnamefont {X.}~\bibnamefont {Zhang}},\ }\bibfield
  {title} {\bibinfo {title} {Observation of vortex-string chiral modes in
  metamaterials},\ }\href {https://doi.org/10.1038/s41467-024-46641-w}
  {\bibfield  {journal} {\bibinfo  {journal} {Nat. Commun.}\ }\textbf {\bibinfo
  {volume} {15}},\ \bibinfo {pages} {2332} (\bibinfo {year}
  {2024})}\BibitemShut {NoStop}%
\end{thebibliography}
\end{document}